\documentclass{jfm}

\usepackage{graphicx}
\usepackage{newtxtext}
\usepackage{newtxmath}
\usepackage{natbib}
\usepackage{hyperref}
\hypersetup{
    colorlinks = true,
    urlcolor   = blue,
    citecolor  = black,
}
\usepackage{placeins}

\newcommand{\RomanNumeralCaps}[1]
\linenumbers

\title{Two-scale interaction of wake and blockage effects in large wind farms}

\author{Andrew Kirby\aff{1}
  \corresp{\email{andrew.kirby@trinity.ox.ac.uk}},
  Takafumi Nishino\aff{1}, \and Thomas D. Dunstan\aff{2}}

\affiliation{\aff{1}Department of Engineering Science, University of Oxford, Parks Road, Oxford OX1 3PJ, UK
\aff{2}Met Office, FitzRoy Road, Exeter EX1 3PB, UK}

\begin{document}
\maketitle

\begin{abstract}
Turbine wake and farm blockage effects may significantly impact the power produced by large wind farms. In this study, we perform Large-Eddy Simulations (LES) of 50 infinitely large offshore wind farms with different turbine layouts and wind directions. The LES results are combined with the two-scale momentum theory (Nishino \& Dunstan 2020, J. Fluid Mech. 894, A2) to investigate the aerodynamic performance of large but finite-sized farms as well. The power of infinitely large farms is found to be a strong function of the array density, whereas the power of large finite-sized farms depends on both the array density and turbine layout. An analytical model derived from the two-scale momentum theory predicts the impact of array density very well for all 50 farms investigated and can therefore be used as an upper limit to farm performance. We also propose a new method to quantify turbine-scale losses (due to turbine-wake interactions) and farm-scale losses (due to the reduction of farm-average wind speed). They both depend on the strength of atmospheric response to the farm, and our results suggest that, for large offshore wind farms, the farm-scale losses are typically more than twice as large as the turbine-scale losses. This is found to be due to a two-scale interaction between turbine wake and \textcolor{blue}{farm induction effects}, explaining why the impact of turbine layout on farm power varies with the strength of atmospheric response.
\end{abstract}

\begin{keywords}
Authors should not enter keywords on the manuscript, as these must be chosen by the author during the online submission process and will then be added during the typesetting process (see \href{https://www.cambridge.org/core/journals/journal-of-fluid-mechanics/information/list-of-keywords}{Keyword PDF} for the full list).  Other classifications will be added at the same time.
\end{keywords}

\section{Introduction}

\par The global wind energy production is set to rise in the next few decades. \textcolor{blue}{To achieve this, wind farm clusters are expected to be built which are an order of magnitude larger than existing ones \citep{Maas2022}.} To optimise their design, it is important to accurately predict the total farm power as well as aerodynamic loads on each turbine.  However, it is very difficult to model the aerodynamics of wind farms because of the multi-scale nature of the wind farm flows \citep{Porte-Agel2020}. In 2019 \O rsted, one of the largest offshore wind farm developers, announced that its wind farms were producing less power than expected \citep{Orsted2019}. The underprediction of farm power was attributed to two effects: wake and farm blockage effects. 

\par Behind every turbine there is a turbulent wake which has a reduced wind speed. When turbine wakes interact with other turbines within a farm this can cause substantial power losses. This is known as the `wake effect' and has been measured to reduce the power of turbines in an existing wind farm by up to 40\% \textcolor{blue}{in the worst case} \citep{Barthelmie2010}. This effect has been extensively investigated in the literature using Large-Eddy Simulations (LES) \citep{Porte-Agel2013,Wu2015,Stevens2016a}. The `farm blockage' is a recently observed effect in large wind farms \citep{Bleeg2018}. The flow resistance caused by a wind farm reduces the wind speed upstream as well inside the farm. Hence, the total power produced by the wind farm is reduced compared to the ideal situation where the upstream wind speed is not affected by the farm \citep{Nishino2020}. 

\par Wind farm aerodynamics have been traditionally modelled using `wake' models, which predict the velocity behind a turbine \citep[e.g.,][]{Jensen1983,Bastankhah2014}. To account for interactions between multiple turbines, the wake velocity deficits are superposed \citep[e.g.,][]{Katic1986,Zong2020}. However, these models do not account for the response of the atmospheric boundary layer (ABL) to the farm. As such they tend to perform poorly for large wind farms \citep{Stevens2016a}. A different approach to modelling wind farms is to use `top-down' models \citep[e.g.,][]{Frandsen1992,Frandsen2006,Calaf2010}. They model the response of an idealised ABL (which \textcolor{blue}{follows a logarithmic law}) to an infinitely large wind farm but cannot take into account the details of turbine-wake interactions explicitly. Hence, these wind farm models cannot correctly capture the two-way interactions of turbine-scale and farm-scale flow effects which determine the performance of large wind farms.

\par \textcolor{blue}{To capture this two-way interaction, it has been proposed to couple wake and top-down models; for example, by adjusting parameters in both models to match the hub-height-averaged velocities \citep{Stevens2016,Starke2021}. Examples of two-way coupling can also be found in existing engineering software, e.g., the `Deep Array Wake Model' \citep{Brower2012}. However, these models involve the coupling of low-order flow models and are therefore limited by the assumptions made by the constituent models, e.g., wake superposition or a log-law wind profile. To account for the effects of more realistic flow physics, it would be beneficial to use an approach based on more fundamental laws of fluid mechanics.}

\par The optimal design of a large wind farm under realistic atmospheric conditions remains a challenge as it requires consideration of the complex atmospheric response to the farm. It is often too expensive to run a large number of simulations which resolve both individual turbines and the atmospheric response to the farm. In addition, to find an optimal design for the long-term performance of a wind farm, such as the annual energy production (AEP), the range of timescales we would need to consider is too wide. As such, \citet{Nishino2020} proposed the `two-scale momentum theory' to split the multi-scale flow problem into `internal' turbine/array-scale and `external' farm/atmospheric-scale problems. \citet{West2020} performed LES of infinitely-large wind farms and the results showed a good agreement with the two-scale momentum theory. However, their LES study was limited to `fully aligned' turbine layouts, and their discussion was also limited to the special case where the momentum supplied by the atmosphere to the wind farm site was fixed. In reality, the strength of atmospheric response to the wind farm resistance depends on mesoscale weather patterns \citep{Patel2021} as well as atmospheric stability and gravity waves \citep[][]{Allaerts2017,Allaerts2018,Allaerts2019}.

\par The aim of the present study is to better understand the fluid mechanics processes which determine the power production of large wind farms, using a combined theoretical and computational approach. First, we will perform a large suite of LES of infinitely large wind farms with different turbine layouts and wind directions. We then combine the results of LES with the two-scale momentum theory to investigate and explain expected performance of large finite-sized wind farms with a realistic range of atmospheric response strengths. Using this approach allows the combined effects of turbine-scale \textcolor{blue}{and farm-scale flow characteristics} on wind farm power to be determined.

\par In section \ref{theory} we summarise the definitions of key wind farm parameters in the two-scale momentum theory \citep{Nishino2020}. Section \ref{les_method} details the methodology of the LES and wind turbine implementation. In section \ref{results} we present the results including validation of the LES code. These results are discussed in section \ref{discussion} and concluding remarks are given in section \ref{conclusions}.

\FloatBarrier

\section{Theory}\label{theory}

\subsection{Two-scale momentum theory}

\begin{figure*}
    \centering
    \includegraphics[width=0.8\textwidth]{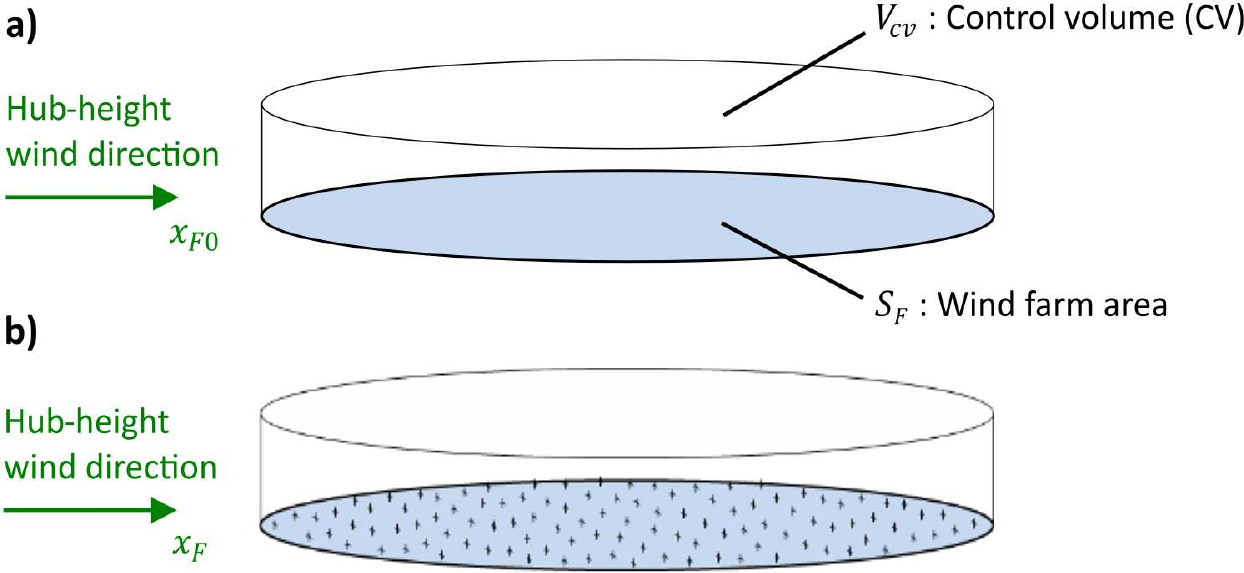}
    \caption{Control volume for an entire wind farm site: a) without turbines and b) with turbines.}
    \label{fig:control_volume}
\end{figure*}

\par Figure \ref{fig:control_volume} shows a pair of control volumes for a given farm site. We first consider the momentum balance of the control volume without the turbines (figure \ref{fig:control_volume}a). For this scenario the following equation can be derived for a `short-time-averaged' flow:

\begin{equation}
    \frac{\partial [\rho U_0]}{\partial t}=X_0-C_0-\left[\frac{\partial{p_0}}{\partial{x_{F0}}}\right]-\frac{\langle\tau_{w0}\rangle S_F}{V_{CV}}
    \label{cv_nofarm}
\end{equation}

\noindent where $U$ is the velocity in the hub-height wind direction (i.e. streamwise direction) $x_F$, $X$ is the net streamwise momentum injection through the top and side boundaries of the control volume (due to advection and Reynolds stress), $C$ is the streamwise component of the Coriolis force averaged over the control volume, $\partial p / \partial{x_F}$ is the pressure gradient in the direction $x_F$, $\tau_w$ is the bottom shear stress, $S_F$ is the wind farm area and $V_{CV}$ is the volume of the control volume. The subscript $0$ refers to values without the turbines present, $[\square]$ refers to control-volume-averaged values and $\langle \square \rangle$ refers to farm-area-averaged values.

\par By considering the control volume with the turbines present (figure \ref{fig:control_volume}b) the following equation can be derived:

\begin{equation}
    \frac{\partial [\rho U]}{\partial t}=X-C-\left[\frac{\partial{p}}{\partial{x_F}}\right]-\frac{\langle\tau_w\rangle S_F+\sum_{i=1}^n T_i}{V_{CV}}
    \label{cv_farm}
\end{equation}

\noindent where $T_i$ is the thrust of turbine $i$ in the farm and $n$ is the total number of turbines in the farm. Equations \ref{cv_nofarm} and \ref{cv_farm} can be combined to obtain the non-dimensional farm momentum (NDFM) equation \citep{Nishino2020}:

\begin{equation}
    C_T^* \frac{\lambda}{C_{f0}} \beta^2 + \beta^\gamma = M
    \label{windfarmmomentum}
\end{equation}

\noindent where $\beta$ is the farm wind-speed reduction factor defined as $\beta\equiv U_F/U_{F0}$ (with $U_F$ defined as the average wind speed in the nominal farm-layer \textcolor{blue}{of height $H_F$}, and $U_{F0}$ is the farm-layer-averaged speed without the presence of the turbines); $\lambda$ is the array density defined as $\lambda\equiv nA/S_F$ (where $A$ is the rotor swept area); $C_T^*$ is the (farm-averaged) `local' or `internal' turbine thrust coefficient defined as $C_T^*\equiv \sum_{i=1}^{n}T_i/\frac{1}{2}\rho U_F^2nA$; $C_{f0}$ is the natural friction coefficient of the surface defined as $C_{f0}\equiv \langle \tau_{w0} \rangle /\frac{1}{2}\rho U_{F0}^2$; $\gamma$ is the bottom friction exponent defined as $\gamma\equiv \log_\beta (\langle \tau_w \rangle/ \langle \tau_{w0} \rangle)$; and $M$ is the momentum availability factor given by:

\begin{equation}
    M = \frac{X-C-\left[\frac{\partial p}{\partial x_F}\right] - \frac{\partial [\rho U]}{\partial t}}{X_0-C_0-\left[\frac{\partial p_0}{\partial x_{F0}}\right] - \frac{\partial [\rho U_0]}{\partial t}}.
    \label{momentumavailability}
\end{equation}

\noindent \textcolor{blue}{Note that the first term in the left-hand-side of  equation \ref{windfarmmomentum} can be extended to include the impact of support structure drag  \citep[see][]{Antoniadis2019} but their impact is usually small and is therefore neglected in this study for simplicity.}

\par The height of the farm-layer, $H_F$, is used to define the reference velocities $U_F$ and $U_{F0}$. $H_F$ is typically between $2H_{hub}$ and $3H_{hub}$ (where $H_{hub}$ is the turbine hub-height); \textcolor{blue}{the NDFM equation (\ref{windfarmmomentum}) is valid as long as the same $H_F$ value is used for both `internal' and `external' problems.} In this study we use a fixed definition of $H_F=2.5H_{hub}$. This is discussed further in appendix \ref{hf_sensitivity}. 

\par The equation \ref{windfarmmomentum} helps the analysis of large wind farm aerodynamics. This is because $C_T^*$ and $\gamma$ on the left hand side are expected to depend primarily on the turbine/array-scale flow physics or `internal' conditions, for example the turbine layout, operating conditions and local wind conditions, whereas $M$ on the right hand side is expected to depend largely on `external' conditions. Following \citet{Nishino2020}, in this study we assume that the `internal' problem (to be modelled using LES in section \ref{les_method} to calculate $C_T^*$ and $\gamma$) can be modelled without explicitly considering the effects of `external' conditions such as wind farm size and location, and the response strength of the atmosphere.

\par The `external' problem is to determine the parameter $M$, which represents how much the amount of momentum available to the farm site differs from its `natural' value. This problem is largely independent of the small-scale flow features and can be modelled using a numerical weather prediction (NWP) model with a wind farm parameterisation, i.e., without resolving individual turbines. \citet{Patel2021} used such an NWP model to demonstrate that, for most cases, $M$ varied almost linearly with $\beta$ (for a realistic range of $\beta$ between 1 and 0.8). Therefore, $M$ can be approximated by 

\begin{equation}
    M=1+\zeta (1-\beta)
    \label{extractability}
\end{equation}

\noindent where $\zeta$ is called the `momentum response' factor or `wind extractability' factor. \citet{Patel2021} found $\zeta$ to vary between 5 and 25 for a typical offshore wind farm site. \textcolor{blue}{Note that $\zeta=0$ corresponds to the case where momentum available to the farm site is assumed to be fixed, i.e., $M=1$. An infinitely-large value of $\zeta$ corresponds to the case where there is no wind speed reduction in the farm-layer, i.e., $\beta=1$. Preliminary results of an extended study from \citet{Patel2021} show that $\zeta$ changes according to atmospheric conditions and decreases exponentially with increasing farm size (see appendix \ref{section:UM_farm_data}).} Although the details of how $\zeta$ changes with weather conditions are still unclear and need to be clarified in future studies, in the present study we take $0<\zeta<25$ as a typical range of the wind extractability for large offshore wind farms. 

\par Using $\beta$ obtained from equation (\ref{windfarmmomentum}) for a set of $C_T^*$, $\gamma$ and $\zeta$, the following equation can be used to calculate the power coefficient of the turbines within the farm,

\begin{equation}
    C_p = \beta^3 C_p^*
    \label{cp}
\end{equation}

\noindent where $C_p$ is the (farm-averaged) turbine power coefficient defined as $C_p\equiv \sum_{i=1}^{n}P_i/\frac{1}{2}\rho U_{F0}^3nA$ ($P_i$ is power of turbine $i$ in the farm) and $C_p^*$ is the (farm-averaged) `local' or `internal' turbine power coefficient defined as $C_p^*\equiv \sum_{i=1}^{n}P_i/\frac{1}{2}\rho U_F^3nA$. 

\subsection{Analytical model of ideal wind farm performance}

\par In this study we consider arrays of actuator discs (or aerodynamically ideal turbines operating below the rated wind speed). For an actuator disc $C_p^*=\alpha C_T^*$ where $\alpha$ is the turbine-scale wind speed reduction factor defined as $\alpha\equiv U_T/U_F$ ($U_T$ is the streamwise velocity averaged over the rotor swept area). We can estimate $\alpha$ using the expression $\alpha=\sqrt{C_T^*/C_T'}$ \textcolor{blue}{where $C_T'\equiv T/\frac{1}{2}\rho U_T^2 A$ is a turbine resistance coefficient describing the turbine operating conditions} (noting that this is strictly valid only for infinitely large regular arrays of turbines where the farm-averaged turbine thrust is identical to the thrust of each individual turbine). The theoretical $C_p$ of an actuator disc is therefore given by

\begin{equation}
    C_p = \beta^3\alpha C_T^* = \beta^3 {C_T^*}^\frac{3}{2}{C_T'}^{-\frac{1}{2}}.
    \label{cp_actuator_disc}
\end{equation}

\noindent \textcolor{blue}{where $C_T^*$ may be predicted using a simple analytical model \citep{Nishino2016} given by}

\begin{equation}
    C_T^* = 4\alpha (1 - \alpha) = \textcolor{blue}{\frac{16C_T'}{(4+C_T')^2}}
    \label{ctstar}
\end{equation}

\noindent \textcolor{blue}{using the expression $C_T'=C_T^*/\alpha^2$ to express $C_T^*$ as a function of $C_T'$. The model predicts $C_T^*$ as a function of turbine-scale wind-speed reduction by using an analogy to the classical actuator disc theory. }This simple analytical model will be compared with LES results later in section \ref{results}. Using the analytical model of $C_T^*$ (equation \ref{ctstar}), and the linear approximation of $M$ (equation \ref{extractability}), equations \ref{windfarmmomentum} and \ref{cp_actuator_disc} can be solved to give a theoretical prediction of \textcolor{blue}{$C_p$, which we will call $C_{p,Nishino}$.} Note that \citet{West2020} also introduced this but only for the special case with $\zeta=0$. \textcolor{blue}{As shown by \citet{Nishino2020}, $C_{p,Nishino}$ is sensitive to $\zeta$ but much less sensitive to $\gamma$.} If we assume that $\gamma=2.0$, then we can obtain an analytical expression for $C_{p,Nishino}$, i.e.,

\begin{equation}
    C_{p,Nishino} = \frac{64C_T'}{(4+C_T')^3} \left[ \frac{-\zeta + \sqrt{\zeta^2 + 4\left( \frac{16C_T'}{(4+C_T')^2}\frac{\lambda}{C_{f0}}+1\right)(1+\zeta)}}{2\left( \frac{16C_T'}{(4+C_T')^2}\frac{\lambda}{C_{f0}}+1\right)} \right]^3.
    \label{cp_nishino}
\end{equation}

\par \textcolor{blue}{It is worth noting that the power coefficient of an isolated turbine, $C_{p,Betz}$, is given by}

\begin{equation}
    \textcolor{blue}{C_{p,Betz} = \frac{64C_T'}{(4+C_T')^3},}
    \label{cp_betz}
\end{equation}

\noindent \textcolor{blue}{which takes the well-known maximum value of $16/27$ at $C_T'=2$. This equation can be obtained by substituting \ref{ctstar} into \ref{cp_actuator_disc} with $\beta=1$ (i.e., assuming flow mechanisms as described by the classical actuator disc theory and no farm-scale wind speed reduction). Note that this is the same as solving equation \ref{cp_nishino} for two special cases: (1) with $\lambda/C_{f0}=0$; and (2) with an infinitely large value of $\zeta$.}

\section{LES modelling}\label{les_method}
\subsection{Governing equations of the flow}\label{LESequations}
We performed LES of flow over periodic turbine arrays using the MetOffice/NERC Cloud (MONC) Model \citep{Brown2018}. The flow is driven by an imposed pressure gradient and is neutrally stratified. The flow is governed by the incompressible Navier-Stokes equations, i.e.,

\begin{equation}
  \frac{\partial{u_i}}{\partial{x_i}}=0
   \label{mass_cont}
\end{equation}

\begin{equation}
  \frac{\partial{u_i}}{\partial{t}}+u_j\frac{\partial{u_i}}{\partial{x_j}} = - \frac{\partial}{\partial{x_i}}\left(\frac{p'}{\rho}\right) + \frac{1}{\rho}\frac{\partial{\tau_{ij}}}{\partial{x_j}} + f_i - \frac{1}{\rho}\frac{\partial{p_{\infty}}}{\partial{x_i}}
     \label{navier_stokes}
\end{equation}

\noindent where $u_i$ is the resolved velocity in the $i$ direction, $p'$ is the pressure perturbation from the reference state, $\rho$ is the reference density, $\tau_{ij}$ is the subgrid stress term, $f_i$ is the force added to model the wind turbines and $\partial{p_{\infty}}/\partial{x_i}$ is the imposed pressure gradient.

The subgrid stress model is a standard Smagorinsky model \citep{SMAGORINSKY1963} given by $\tau_{ij}=\rho \nu S_{ij}$ where $\nu$ is the subgrid-scale eddy viscosity and $S_{ij}$ is the rate of strain tensor. The eddy viscosity is given by a mixing length model $\nu=l^2S$ where $l$ is the mixing length scale and $S$ is the modulus of the rate of strain tensor $S=\|S_{ij}\|/\sqrt{2}$. Near the bottom boundary, the mixing length scale $l$ is damped using the function $1/l^2 = 1/(l_0)^2 + 1/[\kappa (z+z_0)]^2$ described in \citet{Brown1994} where $l_0$ is the basic mixing length scale. $l_0$ is given by $l_0=c_s \Delta$ with a coefficient of $c_s=0.23$ and a \textcolor{blue}{grid spacing given by $\Delta =$ max$(\Delta x,\Delta y)$ \citep{Brown1994}}.

\par \textcolor{blue}{All velocity components are set to zero at the bottom boundary.} \textcolor{red}{The shear stress at the surface is parameterised by specifying $\nu$ following the classical Monin-Obukhov similarity theory.} The horizontal boundary conditions are periodic for all prognostic quantities. The top boundary has a zero vertical velocity boundary condition and a damping layer for the top 200 metres of the domain \textcolor{blue}{(which was not necessary in the present study for neutrally stratified flows but still included for future studies to explore the effect of atmospheric stability).}

\subsection{Actuator disc implementation}\label{LES_actuator_disc}

We model individual turbines as actuator discs following the methodology used by the KULeuven code described in \citet{Calaf2010}. \textcolor{blue}{The approach uses a Gaussian convolution filter to apply the turbine force from the rotor plane onto the LES grid. This allows the position and orientation of turbines to be changed easily.} The thrust force exerted by a single turbine is given by

\begin{equation}
  F = -\frac{1}{2}  \rho C_T' \widehat{U_T}^2 \frac{\pi}{4} D^2
     \label{turbine_force}
\end{equation}

\noindent where $\widehat{U_T}$ is the time-filtered disc-averaged velocity and $D$ is the turbine diameter. This turbine thrust force is spatially distributed using a normalised indicator function $\mathcal{R}(\boldsymbol{x})$, defined as

\begin{equation}
       \mathcal{R}(\boldsymbol{x}) = \frac{4}{\pi D^2} \iint{G(\boldsymbol{x}-\boldsymbol{x'})}\,d^2\boldsymbol{x'}
        \label{indicator function}
\end{equation}

\noindent where $G(\boldsymbol{x})$ is a filtering kernel. This integral is calculated over the surface of the disc. We divide the disc area into 10 segments in the radial and angular directions, respectively. This was sufficient for $\mathcal{R}(\boldsymbol{x})$ to be independent of the number of segments. MONC uses a staggered grid where the $u$ and $v$ velocities are evaluated at different points. As such, two different indicator functions, $\mathcal{R}_x(\boldsymbol{x})$ and $\mathcal{R}_y(\boldsymbol{x})$,  are calculated for the $x$ and $y$ directions. We use the same filtering kernel as described in \citet{Shapiro2019},

\begin{equation}
       G(\boldsymbol{x}) = \left(\frac{6}{\pi \delta^2}\right)^\frac{3}{2} \exp\left(-\frac{6\|\boldsymbol{x}\|}{\delta^2}\right)
       \label{filter kernel}
\end{equation}

\noindent where $\delta$ is the filter width, which following the approach of \citet{Shapiro2019} is given by $\delta = 1.5\sqrt{\Delta x^2 + \Delta y^2 + \Delta z^2}$.

\par The force per unit density at a given grid point $\boldsymbol{x}$ in the direction $i$ is given by

\begin{equation}
       f_i = \frac{1}{2}C_T' \widehat{U_T}^2\frac{\pi}{4} D^2 \mathcal{R}_i(\boldsymbol{x}).
        \label{grid point force}
\end{equation}

The disc-averaged turbine velocity $U_T$ is calculated using the indicator function $\mathcal{R}(\boldsymbol{x})$ as a weighting function, 

\begin{equation}
       U_T = \iiint{u(\boldsymbol{x})\mathcal{R}_x(\boldsymbol{x})\cos{\theta}\,d^3\boldsymbol{x}} + \iiint{v(\boldsymbol{x})\mathcal{R}_y(\boldsymbol{x})\sin{\theta}\,d^3\boldsymbol{x}}
       \label{disc-averaged velocity}
\end{equation}

\noindent where $\theta$ is the wind direction relative to the $x$ direction. \textcolor{blue}{We use a constant value for $\theta$ which is the direction of the pressure gradient forcing.} Note that $u$ refers to the velocity in the $x$ direction whereas $U$ describes velocities in the wind direction.

\par The spatially-averaged velocity $U_T$ is then temporally averaged using a one-sided exponential time filter with a time window of 10 minutes to calculate $\widehat{U_T}$. To calculate $C_T^*$ from the LES we use the following relationship,
\begin{equation}
    C_T^*= \frac{C_T'}{nU_F^2} \sum_{i=1}^n\left(\widehat{U_T}^2\right)_i
\end{equation}

\noindent noting that turbine velocity $U_T$ is time filtered before being squared and then averaged over all $n$ discs. $U_F$ is calculated by integrating the streamwise velocity ($u\cos\theta+v\sin\theta$) between the surface and $2.5H_{hub}$ across the entire domain. Unlike $\widehat{U_T}$, no time filter is used to calculate $U_F$. $C_T^*$ varies with time during the LES so is time-averaged over a long period to give a single value of $\overline{C_T^*}$.

\par To calculate the (farm-averaged) turbine power coefficient from the LES we use the expression,

\begin{equation}\label{les_cp}
    C_p= \frac{C_T'}{nU_{F0}^3} \sum_{i=1}^n\left(\widehat{U_T}^3\right)_i
\end{equation}

\noindent where $U_{F0}$ is the farm-layer-averaged velocity in an LES without turbines. $C_p$ varies with time so $\overline{C_p}$ is calculated by time averaging over a long period.

\FloatBarrier

\section{Results}\label{results}
\subsection{LES code validation}\label{les_validation}

\FloatBarrier

\par We firstly validate \textcolor{blue}{our LES framework with the new actuator disc implementation} by comparing with the benchmark cases reported in \citet{Calaf2010}. We then investigate the sensitivity of our results to horizontal resolution, domain size and pressure solver.

\par
\begin{figure*}
\centering
\includegraphics[width=\textwidth]{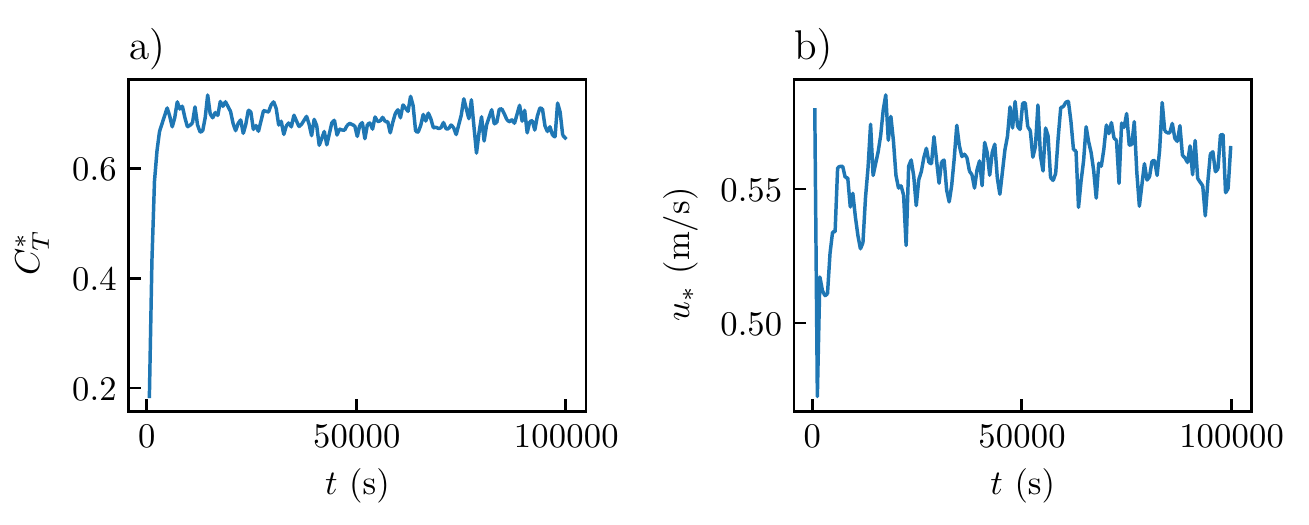}
\caption{Time series of \textcolor{blue}{10-minute-time-averaged} a) internal turbine thrust coefficient $C_T^*$ and b) friction velocity $u_*$ for validation case V-1. \textcolor{red}{Note that the time-averaging here is different from the time-filtering used to calculated $\widehat{U_T}$.}}
\label{fig:LES_convergence}
\end{figure*}

\begin{figure*}
    \centering
    \includegraphics[width=0.9\textwidth]{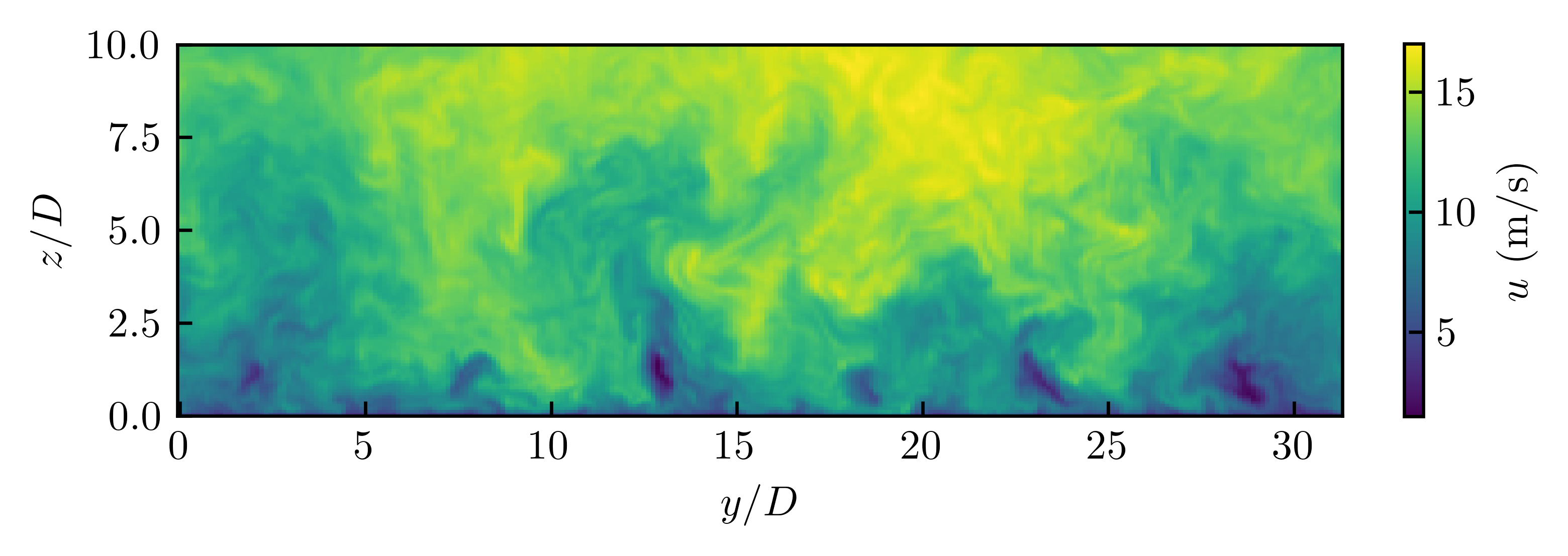}
    \caption{Instantaneous streamwise velocity behind a row of turbines in validation case V-3.}
    \label{fig:valid_inst_yslice}
\end{figure*}

 For the validation cases summarised in table \ref{tab:LES_validation} we use a surface roughness length of $z_0=0.1$m and a pressure gradient of $(1/\rho)$ $dp_{\infty}/dx = 1\times10^{-3} $m/s$^2$. The turbines all have a hub height of 100m and a diameter of 100m. We use a turbine resistance of $C_T'=1.33$ and the same turbine spacing as for Case A1 in \citet{Calaf2010} ($S_x=7.85D$ and $S_y=5.23D$). \textcolor{blue}{The surface roughness length, pressure gradient, turbine design and resistance are chosen to match the values used by \citet{Calaf2010}.} Validation cases V-1, V-2 and V-3 use a FFT pressure solver whereas V-4 uses an iterative pressure solver. Cases V-1, V-3, V-4 have a domain size $L_x \times L_y \times L_z$ of 3.14 $\times$ 3.14 $\times$ 1 km \textcolor{blue}{(with 24 turbines)} and case V-2 has a domain size of  6.28 $\times$ 6.28 $\times$ 1 km \textcolor{blue}{(with 96 turbines)}. All validation cases were run for 100,000 seconds and flow data averaged between $t=$ 30,000 and 100,000 seconds. The convergence of two flow statistics for case V-1 are shown in figure \ref{fig:LES_convergence}. Figure \ref{fig:valid_inst_yslice} shows the instantaneous streamwise velocity plotted on a cross-streamwise plane 2.5$D$ behind a row of turbines in the validation case with a double horizontal resolution. 

\begin{table*}
  \begin{center}
\def~{\hphantom{0}}
  \begin{tabular}{cccccccc}
      Case  & $\Delta x/D$   &   $\Delta y/D$ & $\Delta z/D$ & $L_x \times L_y \times L_z$ (km)& Pressure solver & Uncorrected $\overline{C_T^*}$  & $\overline{C_T^*}$\\[3pt]
       V-1   & 0.245 & 0.245 & 0.0787 & 3.14 $\times$ 3.14 $\times$ 1 & FFT &0.8517 & 0.6845\\
       V-2   & 0.245 & 0.245 & 0.0787 & 6.28 $\times$ 6.28 $\times$ 1 & FFT & 0.8587 & 0.6902\\
       V-3  & 0.1225 & 0.1225 & 0.0787 & 3.14 $\times$ 3.14 $\times$ 1 & FFT & 0.7844 & 0.6957\\
       V-4   & 0.245 & 0.245 & 0.0787 & 3.14 $\times$ 3.14 $\times$ 1 & Iterative & 0.8603 & 0.6914\\
  \end{tabular}
  \caption{Summary of validation cases.}
  \label{tab:LES_validation}
  \end{center}
\end{table*}

\begin{figure*}
\centering
\includegraphics[width=\textwidth]{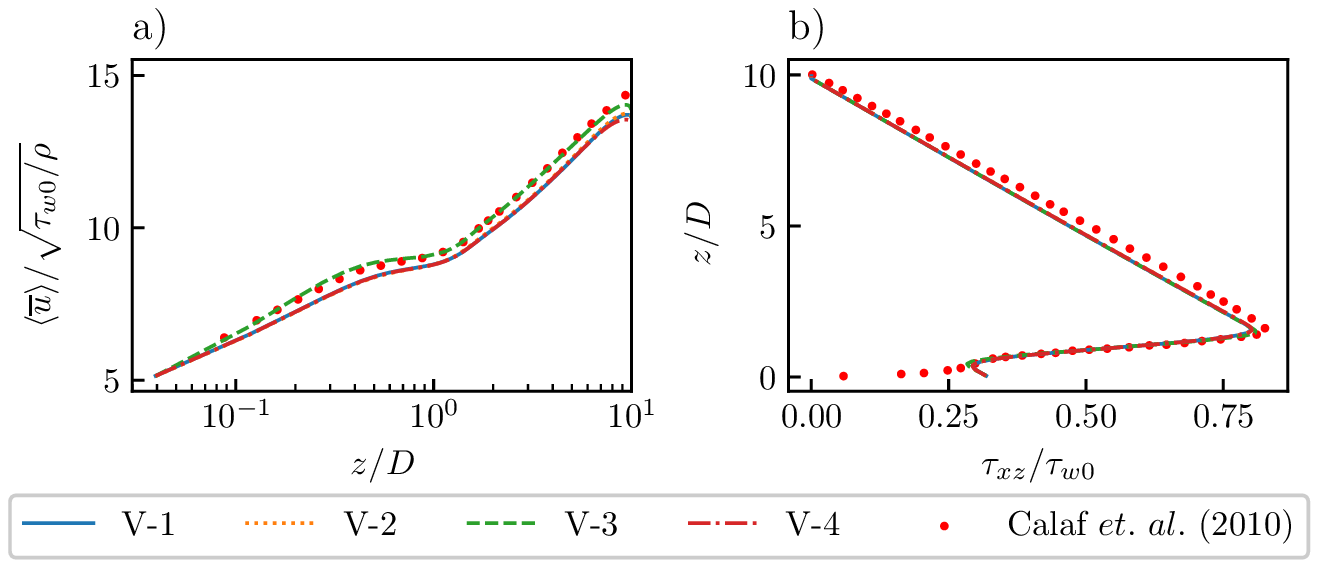}
\caption{a) Mean velocity profiles and b) mean shear stress profiles for all validation cases, compared with Case A1 in \citet{Calaf2010}.}
\label{fig:LES_validation}
\end{figure*}

\par Figure \ref{fig:LES_validation}a shows the time and horizontally averaged streamwise velocity for the four validation cases. The velocity profiles agree well with the results in \citet{Calaf2010}. Figure \ref{fig:LES_validation}b shows the profiles of total shear stress and the results reported in \citet{Calaf2010}. The shear stress profiles match well except for the region near the bottom surface. \textcolor{blue}{Note that the shear stress in our LES does not approach zero at the bottom surface as it includes both modelled and resolved components; the latter of which approaches zero as in the results of \citet{Calaf2010}.} \textcolor{blue}{The velocity profiles are insensitive to the horizontal domain size and the pressure solver.}

\par In wind farm LES using the actuator disc method, the disc-averaged velocity is usually overpredicted \citep{Shapiro2019}. This is because at coarse resolutions, the vorticity shed from the disc edge is not fully captured. This can be seen in our results by considering cases V-1 and V-3 in figure \ref{fig:LES_validation}a. In the coarse grid case V-1, the disc-averaged velocity is overpredicted so the turbine thrust applied is greater. This results in a slightly lower $\langle\overline{u}\rangle$ throughout the entire domain. This effect is also seen in the higher uncorrected $\overline{C_T^*}$ value in case V-1 compared to V-3 (table \ref{tab:LES_validation}), which is due to the overprediction of the disc-averaged velocity.

\par To correct the overprediction of disc velocity in coarse LES the following correction factor was proposed in \citet{Shapiro2019},

\begin{equation}
      N = \left( 1+ \frac{C_T'}{2} \frac{1}{\sqrt{3\pi}}\frac{\delta}{D}\right)^{-1}.
    \label{shapiro_correction}
\end{equation}

\noindent We apply this correction factor to the disc-averaged velocity by multiplying our uncorrected $\overline{C_T^*}$ values by $N^2$. \textcolor{blue}{The correction is applied after the simulation and not during.} \textcolor{blue}{After correction, the horizontal resolution used in cases V-1 and V-3 only had a small impact on $\overline{C_T^*}$, suggesting that this correction factor can be successfully applied to a periodic array of actuator discs.} For the value of $C_T'$ used here the analytical model of $C_T^*$ in (\ref{ctstar}) gives a value of 0.75. All the validation cases in table \ref{tab:LES_validation} have a lower $\overline{C_T^*}$ than this because of wake interactions between turbines.

\begin{figure*}
\centering
\includegraphics[width=\textwidth]{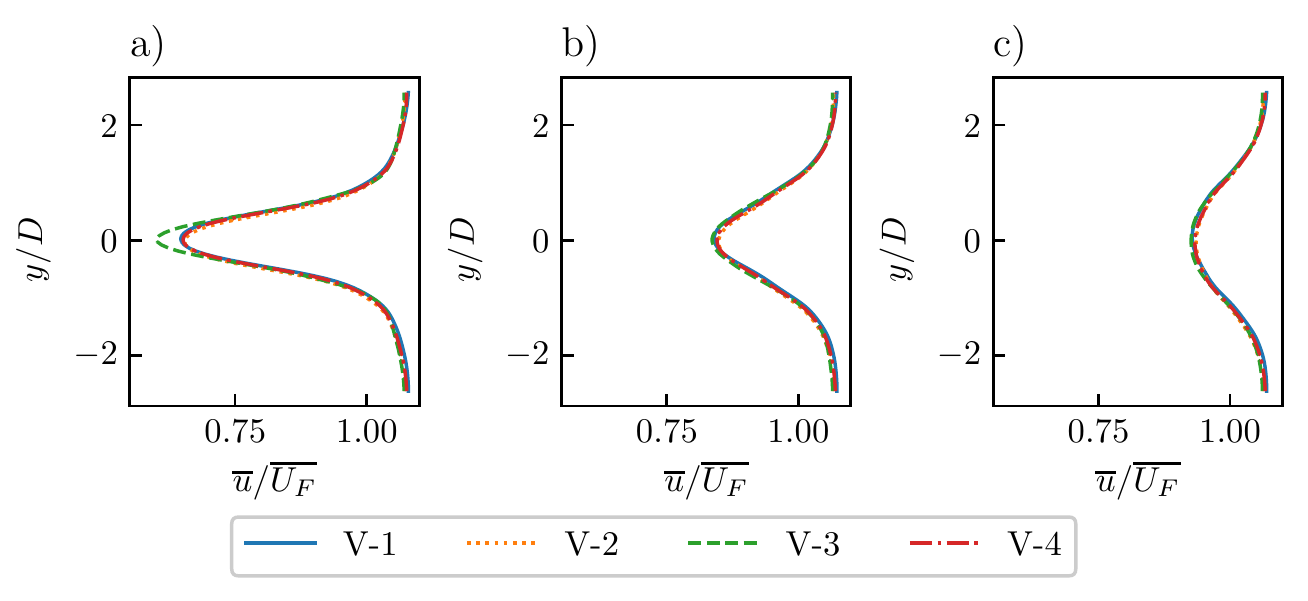}
\caption{Normalised wake velocity deficit for each validation case a) 2D b) 4D and b) 6D downstream.}
\label{fig:wake_resolution}
\end{figure*}

\begin{figure*}
\centering
\includegraphics[width=\textwidth]{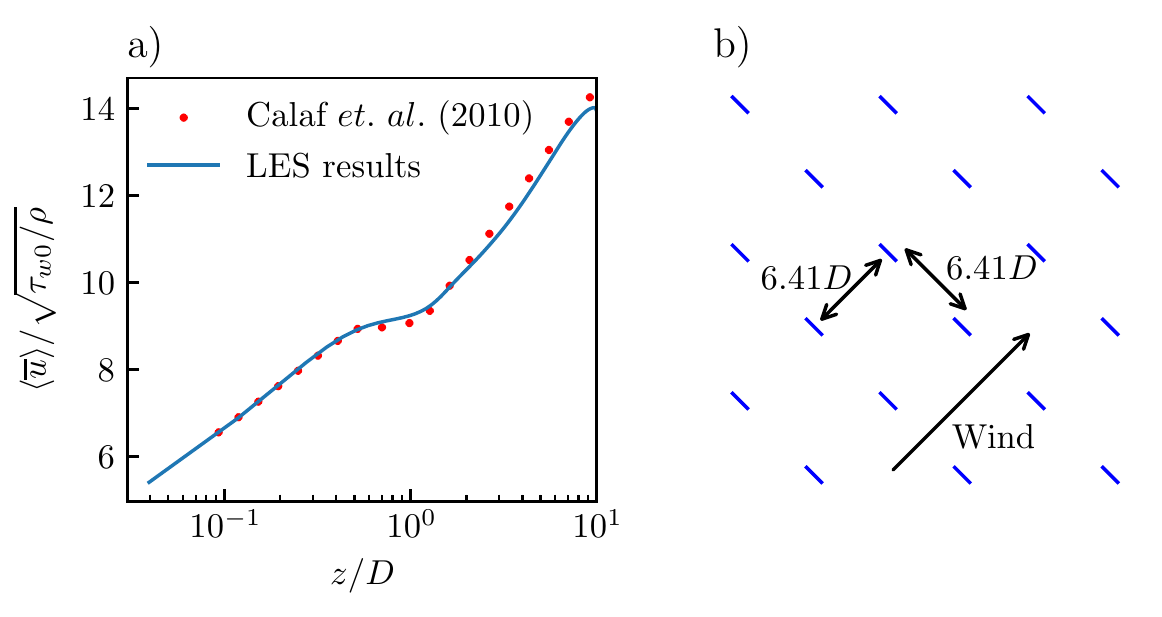}
\caption{Validation case with $45^o$ wind direction: a) mean profile of the horizontally average streamwise velocity and b) turbine layout.}
\label{fig:LES_validation_rotation_u_mean}
\end{figure*}

\par We also consider the effect of resolution on the wake velocity deficit. Figure \ref{fig:wake_resolution} shows the average wake profiles for each of our validation cases in table \ref{tab:LES_validation}. The wake velocity profiles are normalised by the farm-averaged velocity $\overline{U_F}$ for each validation case. \textcolor{blue}{This shows the far wake velocity deficit does not vary with the domain size, horizontal resolution and pressure solver.} Comparing the wake profiles for V-1 and V-3 at 2$D$ downstream (figure \ref{fig:wake_resolution}a) shows a small difference in the velocity deficit at the centre. This is because V-3 uses a different filter size for the projection of the turbine area (see section \ref{LES_actuator_disc}). When comparing the wake profiles 4$D$ and 6$D$ downstream this difference is negligible. This shows that the far wake velocity profile is insensitive to the filter size used for the turbine area projection.

\par To validate the capability of the code to simulate different wind directions, we also performed a simulation with a wind direction of 45$^{\circ}$. The turbine layout in this simulation corresponds to Case K in \citet{Calaf2010} and is shown in figure \ref{fig:LES_validation_rotation_u_mean}b. \textcolor{blue}{We used a resolution of $\Delta x/D \times \Delta y/D \times \Delta z/D$ of $0.227\times0.227\times0.0787$ and a domain size of $L_x \times L_y \times L_z$ of 3.61 $\times$ 3.61 $\times$ 1 km (with 32 turbines).} The horizontally averaged streamwise velocity is shown in figure \ref{fig:LES_validation_rotation_u_mean}a. There is an excellent agreement with the results of \citet{Calaf2010} for the same turbine layout, demonstrating that our new actuator disc implementation for various wind directions (see section \ref{LES_actuator_disc}) is valid.

\FloatBarrier

\subsection{LES results}

\par \textcolor{blue}{The LES cases for this study have the same setup as the validation case V-4 (see section \ref{les_validation}) except for the domain size, the surface roughness length and the streamwise pressure gradient.} To model offshore wind farms, we use a surface roughness length of $1 \times 10^{-4}$m and a pressure gradient of $(1/\rho)$ $dp_{\infty}/dx_F = 8.38 \times 10^{-5}$m/s$^{2}$ in the wind direction $x_F$, which results in $U_{F0}=10.103$m/s and $C_{f0} = 1.607 \times 10^{-3}$ for a fixed nominal farm-layer height of $H_F=250$m (both obtained from LES with no turbines). All cases were run for 100,000 seconds and flow data averaged between $t=$ 60,000 and 100,000 seconds. \textcolor{blue}{Note that we adopted a different spin-up and averaging period (compared to section \ref{les_validation}) because of the different pressure gradient forcing.}

\par
\begin{figure*}
\centering
\includegraphics[width=\textwidth]{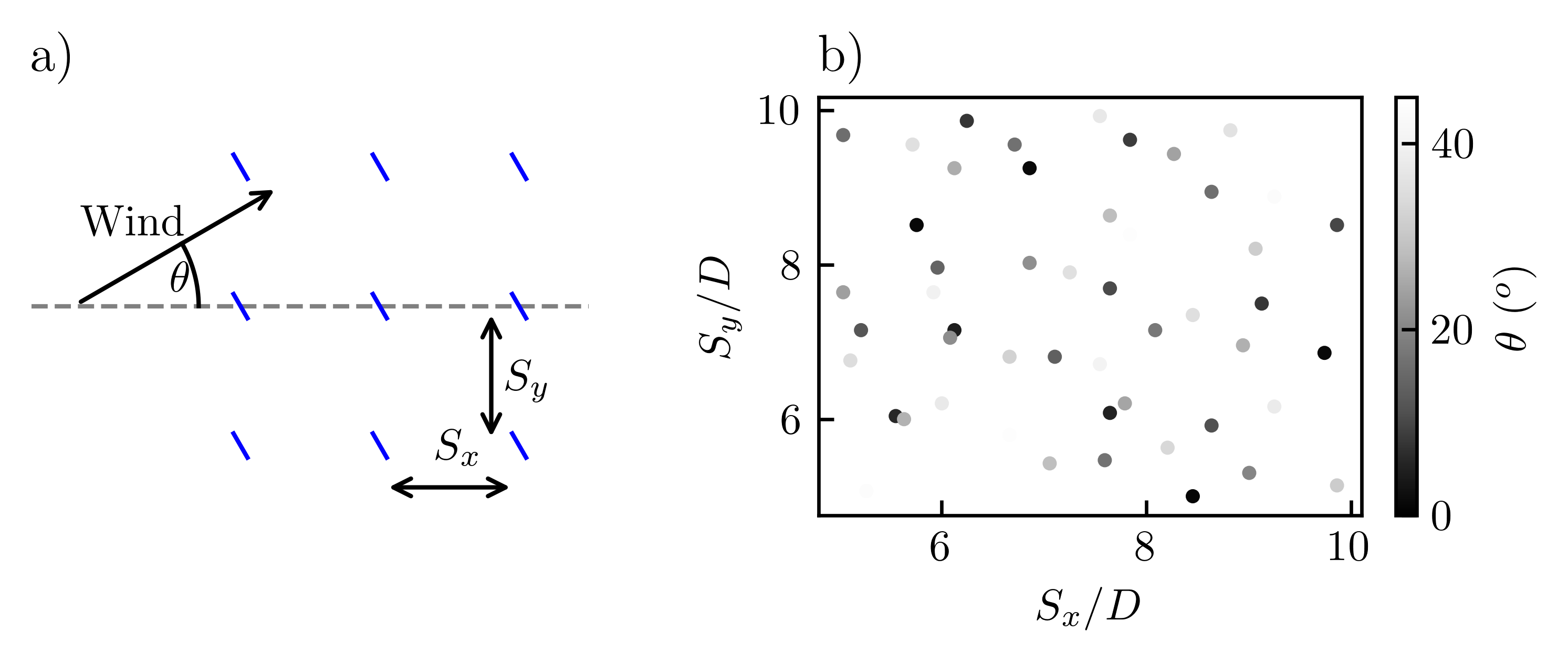}
\caption{Design of numerical experiments: a) input parameters, and b) maximin design of 50 wind farm layouts.}
\label{fig:experiments}
\end{figure*}

 We performed a suite of 50 simulations with different turbine layouts which are described by the parameters $S_x$, the turbine spacing in the $x$ direction, $S_y$, the turbine spacing in the $y$ direction and $\theta$, the wind direction relative to the $x$ direction (see figure \ref{fig:experiments}a). The turbine operating conditions are the same for all simulations and is given by $C_T'=1.33$. We consider a realistic range of turbine layouts and wind directions: $S_x \in [5 D, 10 D]$, $S_y \in [5 D, 10 D]$ and $\theta \in [0^o, 45^o]$. We only consider regular arrays and so by symmetry we only need to consider wind directions up to $45^o$. We adopt the minimum possible horizontal domain size ($L_x$ and $L_y$) within the range between 3.14km and 6.28km (depending on $S_x$ and $S_y$) as the validation results presented in section \ref{les_validation} suggest that the results would be insensitive to the domain size within this range.

\par We use a space filling maximin design \citep{Johnson1990,Santner2018} to select different turbine layouts in the parameter space ($S_x$, $S_y$, $\theta$). The maximin algorithm iteratively selects a point which maximises the minimum distance to other points and to the boundaries of the parameter space. Figure \ref{fig:experiments}b shows the 50 different turbine layouts selected in the parameter space.

\par
\begin{figure}
\centering
\includegraphics[width=\textwidth]{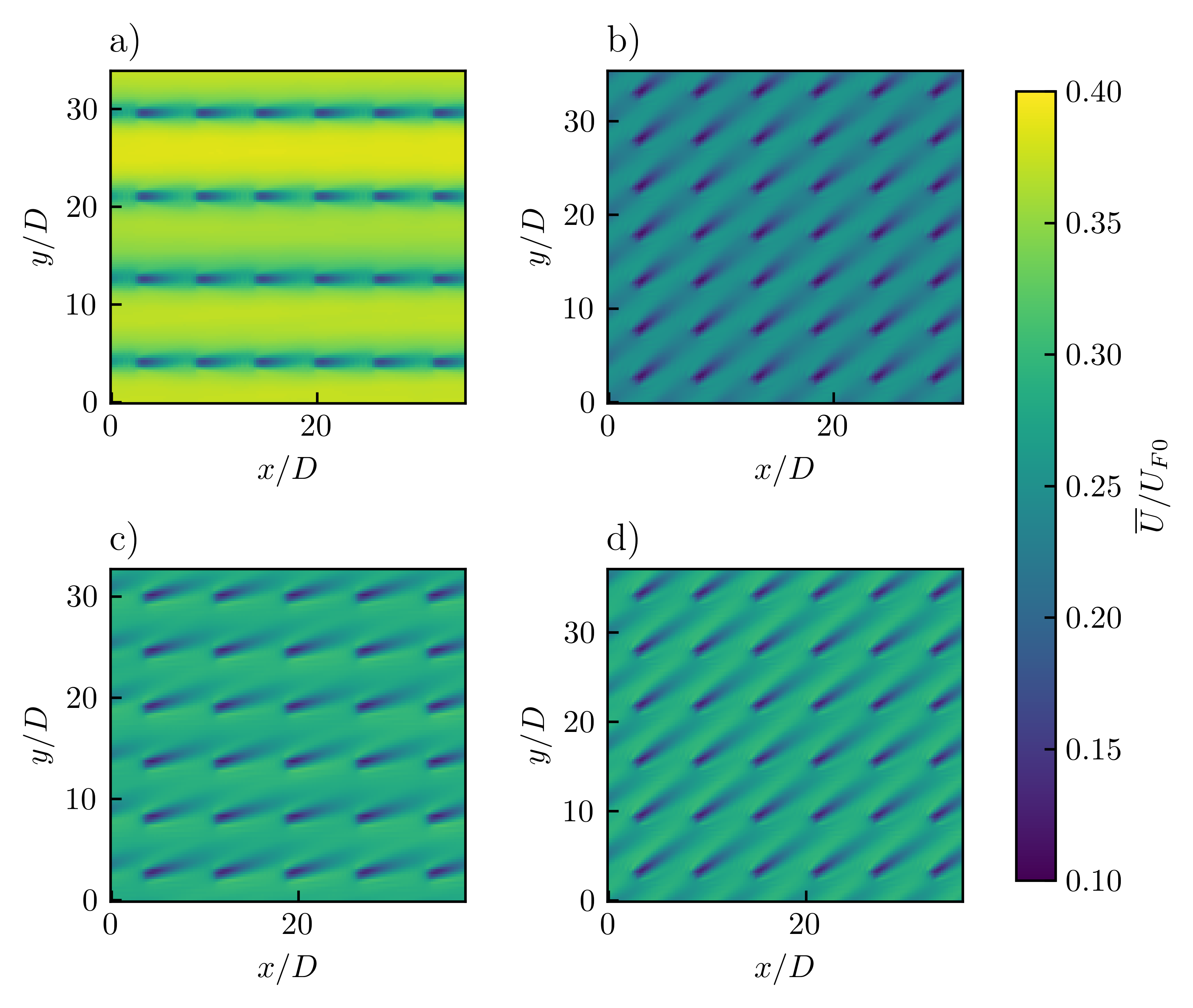}
\caption{Time-averaged streamwise velocity at the turbine hub height for a) $S_x=5.76D, S_y=8.51D, \theta=1.32^o$ b) $S_x=5.27D, S_y=5.07D, \theta=43.8^o$  c) $S_x=7.59D, S_y=5.47D, \theta=16.7^o$  d) $S_x=6.00D, S_y=6.21D, \theta=37.6^o$.}
\label{fig:LES_uwind_hubh}
\end{figure}

 Figure \ref{fig:LES_uwind_hubh} shows the time-averaged flow fields from 4 of 50 cases. Figure \ref{fig:LES_uwind_hubh}a is for a case where the wind direction is almost perfectly aligned with a relatively small streamwise spacing between turbines of 5.76$D$. This case gives a low $\overline{C_T^*}$ value of 0.585 due to strong wake effects. High speed regions between rows of turbines are formed because of the large cross-streamwise turbine spacings and aligned wind direction. Figure \ref{fig:LES_uwind_hubh}b is for a case with a high turbine density and the wind direction almost aligned along the diagonal. This arrangement is similar to a staggered layout. The streamwise spacings between turbines is larger than for figure \ref{fig:LES_uwind_hubh}a so the $\overline{C_T^*}$ has a higher value of 0.669 because of the increased wake recovery between turbines.

\par The flow field for a case with an intermediate wind angle is shown in figure \ref{fig:LES_uwind_hubh}c. The turbine wakes are mostly misaligned with downstream turbines which minimises wake effects and gives a high $\overline{C_T^*}$ value of 0.752. This result agrees qualitatively with \citet{Stevens2014} in which it was found that the maximum farm power was produced by an intermediate wind direction. The results also give further evidence that the analytical model of $C_T^*$ proposed by \citet{Nishino2016} can be used to predict an upper bound to wind farm performance as it gives $C_T^*=0.75$ in this case. Figure \ref{fig:LES_uwind_hubh}d shows the streamwise velocity for a partially waked turbine layout. The partial wake effects cause the $\overline{C_T^*}$ to be reduced slightly to 0.713.

\par Figure \ref{fig:LES_uwind_hubh} also shows the effect of the turbine layout on the farm-averaged wind speed $U_F$. The farm shown in figure \ref{fig:LES_uwind_hubh}a has a low array density and so has a high farm-averaged wind speed \textcolor{blue}{of $\beta=0.347$} (shown by the brighter colour). Figure \ref{fig:LES_uwind_hubh}b shows a farm with a high array density which resulted in a low farm-averaged speed \textcolor{blue}{of $\beta=0.248$.} Figures \ref{fig:LES_uwind_hubh}c and \ref{fig:LES_uwind_hubh}d have similar intermediate array densities and so had intermediate farm-averaged wind speeds \textcolor{blue}{of $\beta=0.289$ and $\beta=0.284$.}

\begin{figure}
\centering
\includegraphics[width=\textwidth]{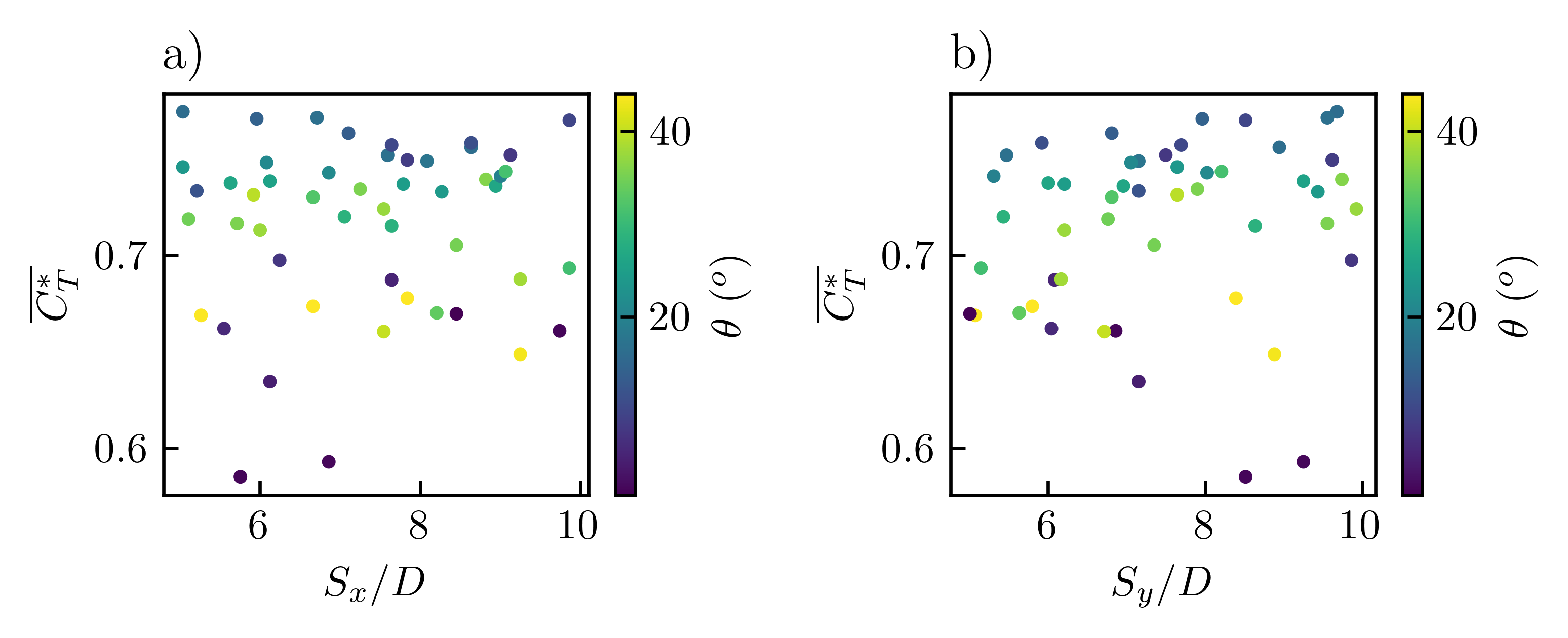}
\caption{$\overline{C_T^*}$ against a) turbine spacing in the $x$ direction $S_x/D$ and b) turbine spacing in the $y$ direction $S_y/D$ with the colour given by the wind direction $\theta$.}
\label{fig:ctstar_sx_sy}
\end{figure}

\begin{figure}
\centering
\includegraphics[width=\textwidth]{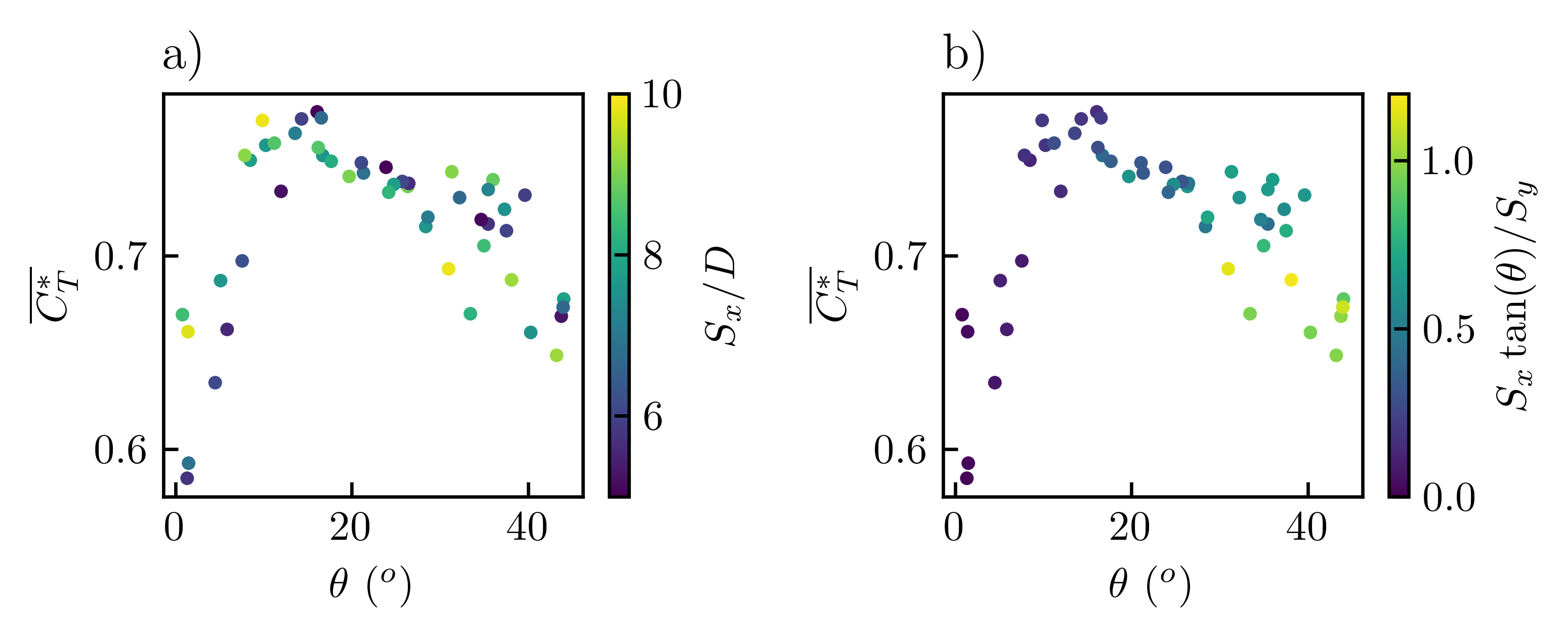}
\caption{$\overline{C_T^*}$ against wind direction with the colour given by a) turbine spacing in the $x$ direction $S_x/D$ and b) $S_x\tan(\theta)/S_y$.}
\label{fig:ctstar_theta}
\end{figure}

\begin{figure*}
\centering
\includegraphics[width=0.5\textwidth]{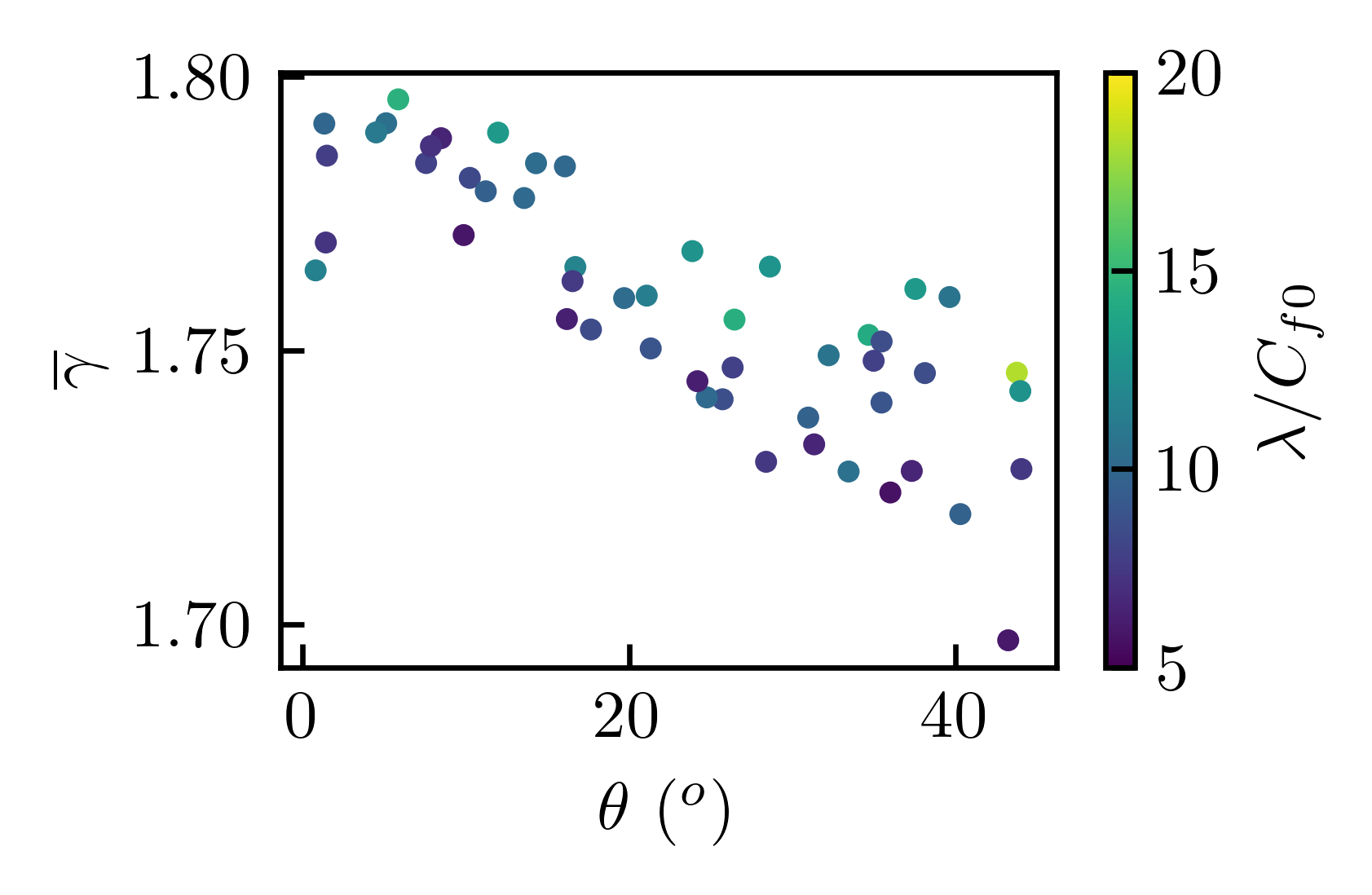}
\caption{$\overline{\gamma}$ against the wind direction $\theta$ with the colour given by the effective array density $\lambda/C_{f0}$.}
\label{fig:gamma}
\end{figure*}

\par Figure \ref{fig:ctstar_sx_sy} shows that $\overline{C_T^*}$ was not a strong function of $S_x$ or $S_y$. $\overline{C_T^*}$ was found to be a much stronger function of $\theta$ (see figure \ref{fig:ctstar_theta}). The lowest $\overline{C_T^*}$ values were for small values of $\theta$ because of the high degree of turbine-wake interactions. When $\theta$ was very small $\overline{C_T^*}$ was also sensitive to the value of $S_x$ (see figure \ref{fig:ctstar_theta}a). As $\theta$ increases $\overline{C_T^*}$ increases rapidly until the maximum value around 15$^o$. As $\theta$ increases further $\overline{C_T^*}$ slowly decreases. This is because turbines start to become aligned along the diagonal (similar to the layout shown in figure \ref{fig:LES_uwind_hubh}b). When $\theta$ is greater than 15$^o$, the minimum $\overline{C_T^*}$ value tends to be observed when $S_x\tan(\theta)/S_y$ is close to 1 (figure \ref{fig:ctstar_theta}b). \textcolor{blue}{This corresponds to layouts where turbines are aligned along the diagonal (similar to the layout shown in figure \ref{fig:LES_uwind_hubh}b).}

\par Figure \ref{fig:gamma} shows that the range of $\overline{\gamma}$ from the wind farm LES is small (varying between 1.7 and 1.8). \citet{Nishino2016} suggested that $\gamma$ would be slightly less than 2 because the presence of turbines would increase the turbulence intensity in the ABL. These results, along with the findings of \cite{Dunstan2018}, provide evidence for this. Figure \ref{fig:gamma} shows that there is a slight variation of $\overline{\gamma}$ with wind direction and effective array density.

\FloatBarrier

\subsection{Prediction of wind farm performance}

\par Now we compare the (farm-averaged) turbine power coefficient $\overline{C_p}$ from the wind farm LES with the analytical model derived from the two-scale momentum theory, $C_{p,Nishino}$ (equation \ref{cp_nishino}).

\par To make a fair comparison of $C_p$ between the theory and the LES, we need to consider the fact that the coarse LES resolution caused the turbine thrust to be overpredicted. For $\overline{C_T^*}$ this was corrected for in post-processing using the correction factor $N$ (equation \ref{shapiro_correction}). However, for $\overline{C_p}$ there are two simultaneous factors that need to be corrected. The first is that the disc velocity (relative to the farm-layer velocity) has been overpredicted, which increases the turbine power. The second is that the farm-layer velocity has been underpredicted (as shown earlier in figure \ref{fig:LES_validation}a), which reduces the turbine power. The first effect can be corrected by using the correction factor $N$ (i.e., multiplying the raw $\overline{C_p}$ from the LES by $N^3$) but the second effect cannot be corrected in this manner.

\par To adjust for the second effect we estimate the farm wind-speed reduction that would be obtained if a sufficiently fine resolution was used for the LES, $\beta_{fine,LES}$. This should be higher than the value obtained from the coarse resolution LES, $\beta_{coarse,LES}$. To calculate $\beta_{fine,LES}$ we assume 

\begin{equation}
      \frac{\beta_{fine,LES}}{\beta_{coarse,LES}} = \frac{\beta_{fine,theory}}{\beta_{coarse,theory}}
    \label{beta_assumption}
\end{equation}

\noindent where $\beta_{fine,theory}$ and $\beta_{coarse,theory}$ are the predictions from the two-scale momentum theory. $\beta_{fine,theory}$ is calculated by solving equation \ref{windfarmmomentum} \textcolor{blue}{(with $M=1$)} using $\overline{C_T^*}$ from the LES and the corresponding array density $\lambda/C_{f0}$. To calculate $\beta_{coarse,theory}$ we do the same but with the uncorrected turbine thrust (i.e.,  $\overline{C_T^*}/N^2$). \textcolor{blue}{Since equation \ref{windfarmmomentum} is derived from the law of momentum conservation, the only assumption we are making in equation \ref{beta_assumption} is that the flow is independent of Reynolds number (which is a reasonable assumption as the change in Reynolds number between the fine and the coarse resolution cases is only of the order of 10\%).} \textcolor{blue}{We can therefore estimate the turbine power from a fine resolution LES, using the expression}

\begin{equation}
\begin{split}
      \textcolor{blue}{\frac{\overline{C_p}_{fine,LES}}{\overline{C_p}_{coarse,LES}} = \left( \frac{{U_T}_{fine,LES}}{{U_T}_{coarse,LES}} \right)^3 = \left( \frac{{U_T}_{fine,LES}/{U_F}_{fine,LES}}{{U_T}_{coarse,LES}/{U_F}_{coarse,LES}} \right)^3 \left( \frac{{U_F}_{fine,LES}}{{U_F}_{coarse,LES}}\right)^3} \\ \textcolor{blue}{=N^3 \left(\frac{\beta_{fine,LES}}{\beta_{coarse,LES}} \right)^3}
    \label{cp_adjustment}
\end{split}
\end{equation}

\noindent where $\overline{C_p}_{coarse,LES}$ is from the coarse resolution wind farm LES (equation \ref{les_cp}).

\begin{figure*}
\centering
\includegraphics[width=\textwidth]{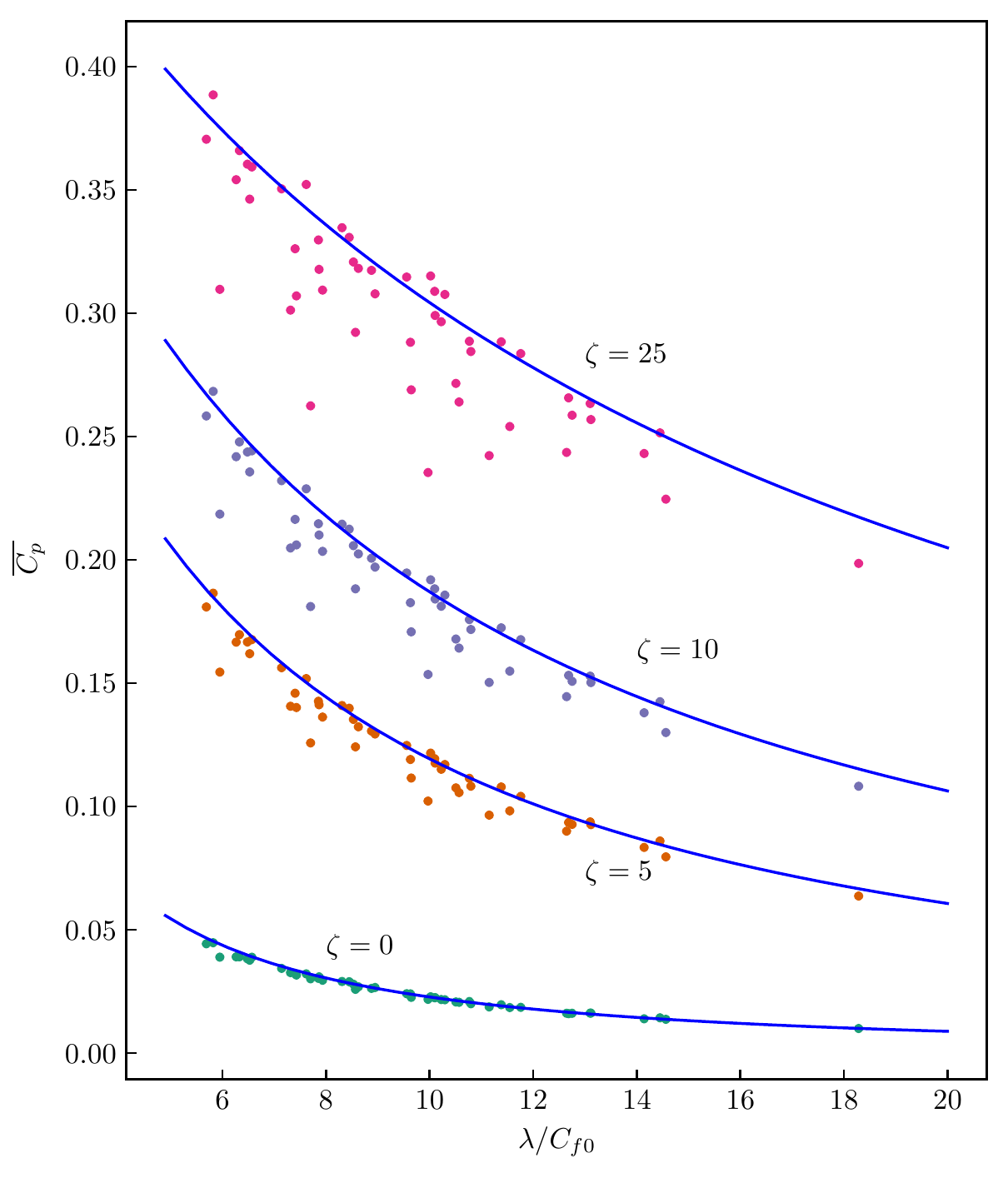}
\caption{Average turbine power coefficient $\overline{C_p}$ from the 50 wind farm LES adjusted for different wind extractability factors ($\zeta$). The $C_p$ predicted by the two-scale momentum theory (equation \ref{cp_nishino} with $C_T'=1.33$) is shown by the blue lines for comparison.}
\label{fig:LES_cp_zeta}
\end{figure*}

\par The green symbols marked by $\zeta$=0 in figure \ref{fig:LES_cp_zeta} show the corrected $\overline{C_p}$ for each of the 50 wind farm LES runs. These are plotted against the effective array density $\lambda/C_{f0}$. The blue line shows the $C_p$ prediction using the theoretical model (equation \ref{cp_nishino}). The theory matches remarkably well with the LES results, despite that the theory does not account for the turbine layout parameters, namely $S_x$, $S_y$ and $\theta$. The reason for this excellent agreement for $\zeta=0$ will be discussed later in section \ref{discussion}. 

\par Next, we use the results from LES of infinitely large wind farms to estimate the average power coefficients for large but finite-sized wind farms, following the concept of the two-scale momentum theory. \textcolor{blue}{We combine the infinite wind farm LES results with the simple linear model of the momentum availability factor $M$ (equation \ref{extractability}). This does not capture the finite-size effects observed near the edge of the farm but reveals general trends of turbine layout effects with different atmospheric responses. We assume that the farms are still sufficiently large that the flow over the farm is mostly fully developed (or more specifically, the dependency of $C_T^*$ on $S_x$, $S_y$ and $\theta$ is still approximately the same as that in the corresponding infinitely-large farm). `Sufficiently large' depends on atmospheric conditions \citep{Wu2017} but it is likely to be on the order of 10km.} 

\par Firstly, we consider the balance of the pressure gradient forcing (PGF) with surface stress and turbine thrust \textcolor{blue}{in our LES},

\begin{equation}
      \rho (S_FC_f^*+nAC_T^*)U_F^2/2 = S_F L_z \Delta p/\Delta x
    \label{pgf_balance}
\end{equation}

\noindent where $C_f^*$ is the `local' or `internal' friction coefficient $C_{f}^*\equiv \langle \tau_{w} \rangle /\frac{1}{2}\rho U_{F}^2$ and $\Delta p/\Delta x$ is the PGF applied to the LES of an infinitely large wind farm.  

\par If we assume $C_f^*$ and $C_T^*$ are independent of the PGF (as the Reynolds number of the flow is very high), then equation \ref{pgf_balance} shows that $U_F^2$ should be proportional to $\Delta p / \Delta x$; hence
\begin{equation}
      \frac{U_{F,finite}^2}{U_F^2} = \frac{\Delta p_{finite}}{\Delta x} / \frac{\Delta p}{\Delta x}
    \label{zeta_UF_increase}
\end{equation}

\noindent where $U_F$ is the farm-layer-averaged wind speed in the LES and $\Delta p / \Delta x$ is the corresponding PGF, $U_{F,finite}$ is the new farm-layer-averaged wind speed that would result from an unknown PGF for a finite-sized wind farm, $\Delta p_{finite} / \Delta x$.

\par As the flow is quasi-steady and horizontally periodic with no Coriolis force in this study \textcolor{blue}{(meaning that we ignore the contributions of all terms except the PGF in equation \ref{momentumavailability})}, the right hand side of equation \ref{zeta_UF_increase} is equivalent to the momentum availability factor $M$ in equation \ref{momentumavailability}. \textcolor{blue}{This will not be true for real wind farms under real atmospheric conditions. However, this is valid for our simplified analysis where the flow across the farm is mostly fully developed and is driven purely by a PGF.} As noted earlier, it was found in \citet{Patel2021} that $M$ can be well approximated by $M=1+\zeta (1-\beta)$ and $\zeta$ was typically between 5 and 25 for an offshore wind farm site. Substituting this expression for $M$ into equation \ref{zeta_UF_increase} gives

\begin{equation}
      \frac{U_{F,finite}^2}{U_F^2} = 1 + \zeta \left(1-\frac{U_{F,finite}}{U_{F0}}\right).
    \label{zeta_UFprime_quadratic}
\end{equation}

\noindent Since $U_F$ and $U_{F0}$ are both known from LES results, $U_{F,finite}$ can be calculated analytically for a given $\zeta$. Finally we assume (again based on the Reynolds number independency) that the disc-averaged velocity $U_T$ scales with $U_F$ and as such 

\begin{equation}
      C_{p,finite}=C_p \left(\frac{U_{F,finite}}{U_F}\right)^3.
    \label{zeta_Cp_increase}
\end{equation}

\noindent where $C_{p,finite}$ is the average turbine power coefficient of the large finite-sized wind farm.

\par Figure \ref{fig:LES_cp_zeta} also shows the average power coefficients $\overline{C_p}$ estimated for three different $\zeta$ values. This shows that the atmospheric response can significantly impact the farm power. $\overline{C_p}$ for $\zeta=25$ are roughly an order of magnitude higher than $\overline{C_p}$ for the same layout with $\zeta=0$. Note that these results are with a constant turbine operating condition of $C_T'=1.33$ which may not be optimal for a given $\zeta$. Figure 4 in \cite{Nishino2020} shows how the theoretically optimal turbine power coefficient, $C_{p,max}$ varies with the strength of atmospheric response. 

\begin{figure*}
\centering
\includegraphics[width=\textwidth]{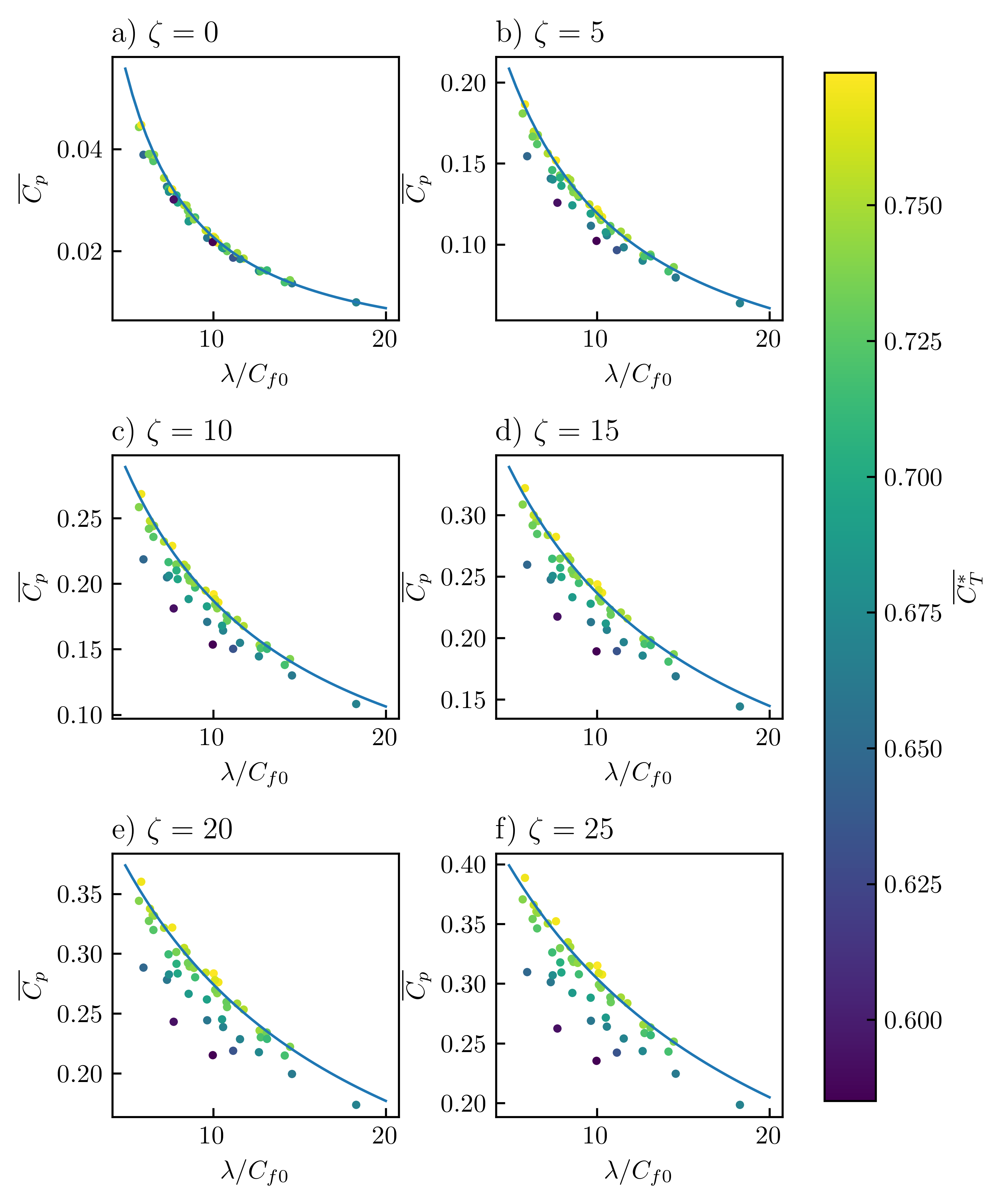}
\caption{Turbine power coefficient $\overline{C_p}$ for a) infinite wind farms ($\zeta=0$) and b) to f) finite wind farms with the colour giving the $\overline{C_T^*}$ recorded for each layout. The $C_p$ predicted by the two-scale momentum theory is shown by the blue line.}
\label{fig:LES_cp}
\end{figure*}

\par Of particular interest in figure \ref{fig:LES_cp_zeta} is that, for a given $\lambda/C_{f0}$, the variation of $\overline{C_p}$ (due to different turbine layouts) increases with $\zeta$. To better understand this trend, the $\overline{C_p}$ values for the 50 turbine layouts are presented together with their $\overline{C_T^*}$ values in figure \ref{fig:LES_cp} for 6 different $\zeta$ values separately. Figure \ref{fig:LES_cp}a shows the results for infinitely large farms with $\zeta$=0. This shows that the average turbine power coefficient is a strong function of the effective array density and insensitive to $\overline{C_T^*}$ meaning that turbine-wake interactions have little effect on the farm power. However, for finite-sized farms (figures \ref{fig:LES_cp}b-f) the layouts with high wake interactions and a low $\overline{C_T^*}$ value tend to produce less power. These are shown by the darker plots which fall well below the theoretical prediction (blue line). Interestingly, some of the layouts seem to produce slightly higher power than predicted by the two-scale momentum theory. These layouts have a $\overline{C_T^*}$ slightly higher than 0.75 (the value for an isolated turbine) and this seems to be due to locally accelerated flow caused by the local blockage effect \citep{Nishino2015,Ouro2021}. These results suggest that both the array density and turbine-wake interactions are important for the performance for large finite-sized wind farms. 

\par Figure \ref{fig:LES_cp} also suggests that finite-sized wind farms are less sensitive to the effective array density than infinitely-large farms. Figure \ref{fig:LES_cp}a shows a roughly inverse relationship between $\overline{C_p}$ and $\lambda/C_{f0}$ whereas figures \ref{fig:LES_cp}b-f show a more linear decrease of $\overline{C_p}$. Overall, the \textcolor{blue}{analytical model (equation \ref{cp_nishino})} predicts the variation of $\overline{C_p}$ with the effective array density well for all of the atmospheric responses. The theoretical model does not predict the effect of turbine-wake interactions because the model of $C_T^*$ (equation \ref{ctstar}) is a function of turbine operating conditions only. These results support the argument that the two-scale momentum theory can be used to provide an \textcolor{blue}{approximate} upper limit on wind farm performance.

\FloatBarrier

\section{Discussion}\label{discussion}

\par \textcolor{blue}{The analytical model of $C_T^*$ (equation \ref{ctstar}) has been shown in this study to provide an approximate upper bound to wind farm performance. The model is based on the classical actuator disc theory with an upstream velocity of $U_F$. Equation \ref{ctstar} would provide accurate predictions of $C_T^*$ if the following two conditions are met: (1) the wind speed upstream of each turbine in the farm is $U_F$; and (2) the mechanism of the flow around each turbine is the same as that around an isolated actuator disc. In reality, the wind speed upstream of each turbine is often lower than $U_F$ due to wake effects, reducing $C_T^*$ (and thus $C_p$). Conversely, local blockage effects \citep{Nishino2015} may increase $C_T^*$ because (1) it creates locally accelerated flows which may allow the upstream velocity of most turbines to be higher than $U_F$ \citep[see e.g.,][]{Ouro2021}; and (2) it changes the mechanism of the flow around each turbine, allowing for a higher $U_T$ (for a given $C_T'$) than that predicted by the classical actuator disc theory. However, such a positive effect of local blockage can be exploited only when the layout is carefully optimised for a specific wind direction. As such, equation \ref{ctstar} can be used to predict the upper bound of farm performance, or the performance of an ideal wind farm without the negative effect of turbine-wake interactions.}

\par The results from this study also show that the $\overline{C_p}$ of infinitely large wind farms depends mainly on the effective array density (see figure \ref{fig:LES_cp}a), i.e., the farm power is insensitive to the turbine-scale flow interactions. A closer look at the results suggest that the limited impact of turbine-scale interactions is due to the fact that the turbine drag is typically much greater than the surface drag. The ratio of total turbine drag to surface drag is given by:

\begin{equation}
    \frac{\sum_{i=1}^n T_i}{\langle \tau_w \rangle S_F}=\frac{\frac{1}{2}\rho C_T^* n A U_F^2}{\frac{1}{2}\rho C_f^* S_F U_F^2} = \frac{\lambda C_T^*}{C_f^*}.
    \label{ratioturbinesurfacedrag}
\end{equation}

\begin{figure*}
\centering
\includegraphics[width=\textwidth]{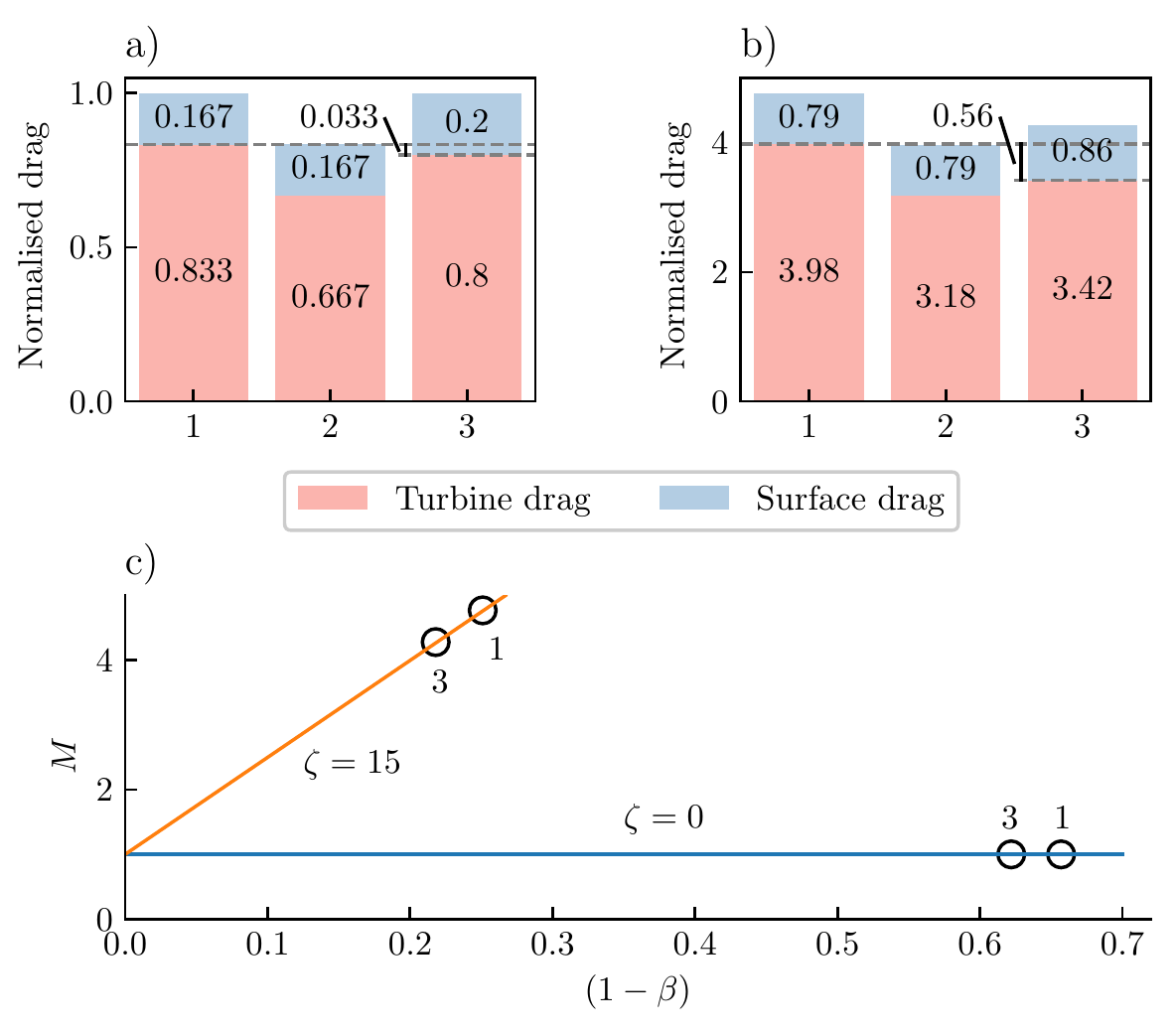}
\caption{Examples of offshore wind farm drag composition at 3 different states: 1) equilibrium state before wind direction change, 2) non-equilibrium state immediately after a small change of wind direction, and 3) new equilibrium state after atmospheric response, for a) $\zeta=0$ and b) $\zeta=15$. c) schematic of the three states on the $M$ vs $(1-\beta) $ plot.}
\label{fig:drag_composition}
\end{figure*}

\noindent We measured the time-averaged value of $\lambda C_T^*/C_f^*$ for all 50 turbine layouts simulated. The mean value was 5.22, the minimum value was $2.93$ and the maximum was $8.59$. Therefore, the turbine drag was typically 5 times greater than the surface drag. Consider an offshore farm where $\lambda C_T^*/C_f^*=5$, $\lambda /C_{f0}=10$ (both typical values) and $\zeta=0$. This means the turbine drag is 5 times greater than the surface drag and the momentum supplied by the atmosphere does not change in response to the farm. The composition of the total drag (normalised by $\langle \tau_{w0} \rangle S_F$) for this scenario is shown by the bar 1 in figure \ref{fig:drag_composition}a. This state 1 is an equilibrium state, i.e., the normalised drag is balanced by $M$ as in equation \ref{windfarmmomentum}. If there is a sudden small change in wind direction which increases the degree of wake interactions then this will decrease $C_T^*$ (or the ratio of $T$ to $U_F^2$) for the farm. In this example, the turbine drag is now only 4 times greater than the surface drag (bar 2 in figure \ref{fig:drag_composition}a). This corresponds to $C_T^*$ decreasing from $0.75$ to $0.6$ (which is close to the largest difference observed in the 50 wind farm LES). This state 2 is a non-equilibrium state, i.e., the normalised drag is not balanced by $M$ (due to the sudden small change of wind direction). However, if the momentum supplied by the atmosphere is unchanged the wind speed in the farm $U_F$ will eventually increase to compensate for the reduced turbine drag. The surface and turbine drag both scale with $U_F^2$ so they are both expected to increase at the same rate. Bar 3 in figure \ref{fig:drag_composition}a shows the new equilibrium state with a new composition of turbine and surface drag after the increase in $U_F$. Comparing states 1 and 3, the total turbine drag has been reduced only slightly because of the constant amount of momentum supplied by the atmosphere to the farm site. Therefore, for aerodynamically ideal turbines (or turbines operating below the rated wind speed), the average turbine power coefficient $C_p$ is also insensitive to turbine-scale flow interactions when the momentum supplied by the atmosphere is constant ($\zeta=0$).

Figure \ref{fig:drag_composition}b explains why turbine-scale flow interactions become more important as $\zeta$ increases. Now the momentum supplied by the atmosphere changes with the farm wind-speed reduction factor $\beta$ according to $M=1+\zeta(1-\beta)$. The ratios between turbine and surface drag for states 1 and 2 are exactly the same as for the $\zeta=0$ case in figure \ref{fig:drag_composition}a. However, now when the wind speed increases in response to the reduced turbine drag the momentum supplied by the atmosphere to the farm site changes. As the wind speed $U_F$ increases, $\beta$ increases so the momentum supplied by the atmosphere decreases (see figure \ref{fig:drag_composition}c). Therefore the sum of the turbine and surface drag for state 3 decreases compared to state 1. This explains why there is a much greater reduction in turbine drag (and thus power) for the non-zero $\zeta$ case than for $\zeta$=0. As $\zeta$ increases there will be a larger decrease in the total drag in response to an increased wind speed. Therefore the power losses due to turbine-scale flow interactions also increase as $\zeta$ increases.

\par To better understand the factors which determine the power output of wind farms we propose three power loss factors. Firstly, the turbine-scale loss factor $\Pi_T$ which is defined by

\begin{equation}
    \Pi_T \equiv 1 - \frac{C_p}{C_{p,Nishino}}
    \label{turbine_loss_factor}
\end{equation}

\begin{figure*}
\centering
\includegraphics[width=\textwidth]{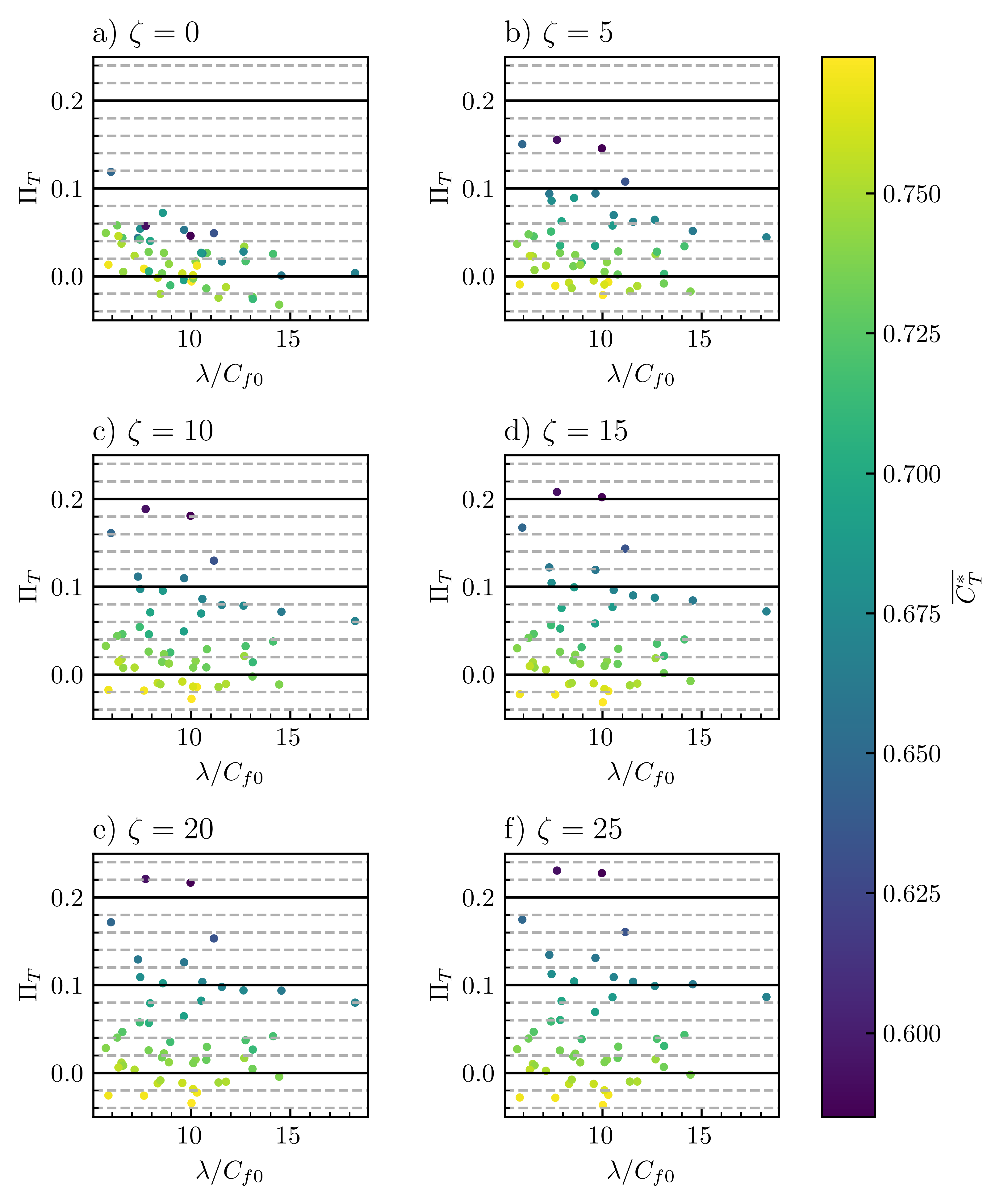}
\caption{Turbine-scale loss factor $\Pi_T$ for a) infinite wind farms ($\zeta=0$) and b) to f) finite wind farms with the colour giving the $\overline{C_T^*}$ recorded for each layout.}
\label{fig:LES_turbinelossfactor}
\end{figure*}

 \noindent \textcolor{blue}{where $C_{p,Nishino}$ is given by equation \ref{cp_nishino}.} $\Pi_T$ represents the power losses due to turbine-scale (or internal) flow interactions only, separate from the losses due to the farm-scale atmospheric response. For real turbines, $\Pi_T$ would also include turbine design losses (i.e., power losses due to a non-ideal rotor design for a given $C_T'$). Figure \ref{fig:LES_turbinelossfactor}a shows that the losses caused by turbine interactions are small for infinitely large farms, typically less than 5\%. \textcolor{blue}{For some turbine layouts $\Pi_T$ is negative, meaning that $C_p$ exceeds $C_{p,Nishino}$. These layouts have $\overline{C_T^*}$ values greater than 0.75 which is the value given by equation \ref{ctstar}. As discussed earlier, this is likely to be due to local blockage effects increasing the turbine incident velocity above the farm-layer velocity $U_F$.} Figures \ref{fig:LES_turbinelossfactor}b-f show that the turbine-scale losses are greater in finite-sized farms with the same turbine layout. Under different atmospheric conditions, the same turbine layout can give different turbine-scale losses. Across a realistic range of wind extractability factors, the turbine-scale losses from the same layout can vary significantly. As an example, the losses from one layout varies from 13\% to 22\% as $\zeta$ changes from 5 to 25. Figure \ref{fig:LES_turbinelossfactor} suggests the maximum losses due to turbine-scale flow interactions is likely to be about 20\% for large offshore wind farms.

\par The farm-scale loss factor, $\Pi_F$, represents the power loss due to the atmospheric response to the whole farm, and is defined by 

\begin{equation}
    \Pi_F \equiv 1 - \frac{C_{p,Nishino}}{C_{p,Betz}}
    \label{farm_loss_factor}
\end{equation}

\begin{figure*}
\centering
\includegraphics[width=\textwidth]{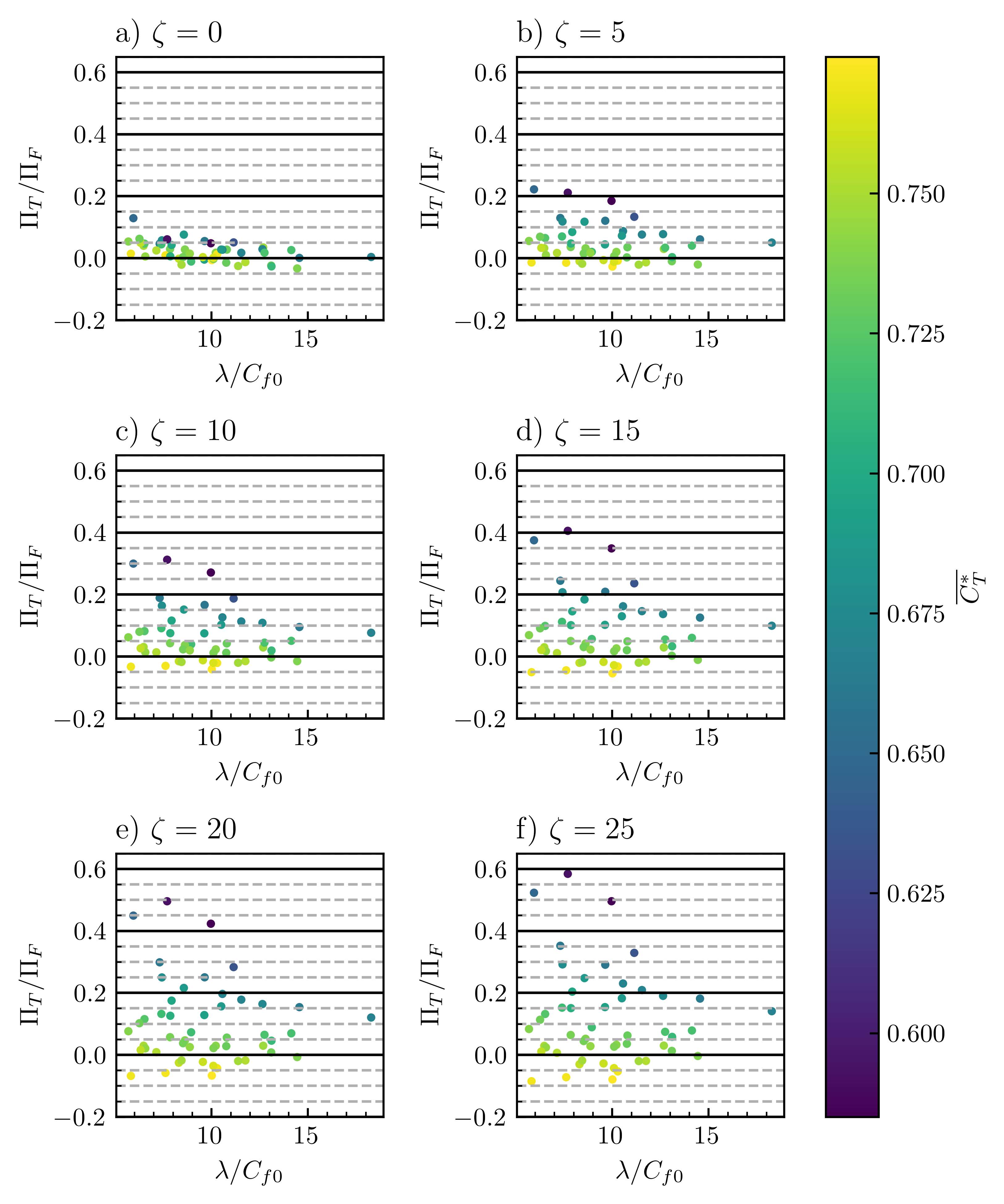}
\caption{Ratio of turbine-scale to farm-scale loss factors $\Pi_T/\Pi_F$ for a) infinite wind farms ($\zeta=0$) and b) to f) finite wind farms with the colour giving the $\overline{C_T^*}$ recorded for each layout.}
\label{fig:LES_lossfactorratio}
\end{figure*}

\noindent \textcolor{blue}{where $C_{p,Betz}$ is given by equation \ref{cp_betz}. Note that in this study $C_T'=1.33$ and hence $C_{p,Betz}=0.563$ ($C_T'=2$ would give the optimal performance for an isolated turbine of $C_{p,Betz}=16/27$)}. $\Pi_F$ represents the power loss accompanied by the reduction of the farm-average wind speed. Figure \ref{fig:LES_lossfactorratio} shows that the farm-scale losses are typically more than twice the turbine-scale losses (i.e.,  $\Pi_T/\Pi_F$ is generally less than 0.5). This suggests that for large offshore wind farms the atmospheric response to the array density is more important than the turbine-scale interactions (i.e., wake effects). Similarly the total power loss factor, $\Pi$, can also be defined as 

\begin{equation}
    \Pi \equiv 1 - \frac{C_p}{C_{p,Betz}} = 1 - (1-\Pi_T)(1-\Pi_F).
    \label{total_loss_factor}
\end{equation}

\par
\begin{figure*}
\centering
\includegraphics[width=\textwidth]{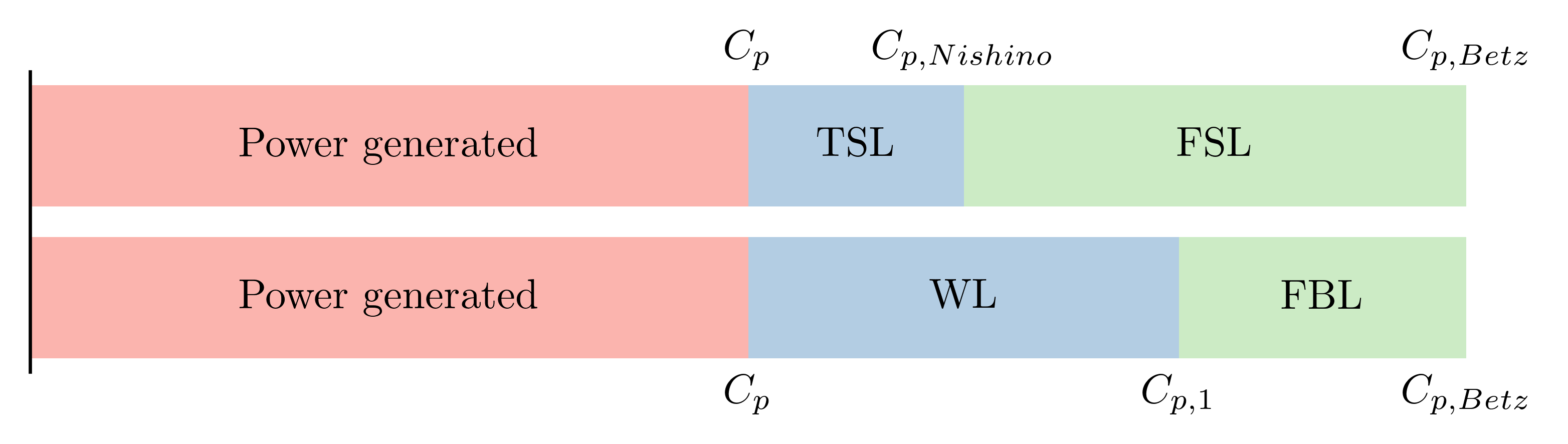}
\caption{Comparison of turbine-scale loss (TSL) and farm-scale loss (FSL) with what is known as wake loss (WL) and farm blockage loss (FBL). $C_{p,1}$ is the power coefficient recorded by the first row of turbines in a farm.}
\label{fig:loss_factors}
\end{figure*}

 It is important to note that the turbine-scale loss factor discussed above is smaller than what is typically referred to as `wake losses' (see figure \ref{fig:loss_factors}). `Wake losses' are traditionally evaluated by comparing the farm power with the power produced by the first row of turbines ($C_{p,1}$ in figure \ref{fig:loss_factors}). However, this includes not only the losses due to wake interactions between turbines but also the atmospheric response to the array density. In contrast, $\Pi_T$ represents the power losses solely due to the interactions between turbines. For the same reason, the farm-scale loss factor, $\Pi_F$, is larger than what is often referred to as `farm blockage losses'. These two different classifications of power losses (one using $C_{p,Nishino}$ and the other using $C_{p,1}$ as a point of reference) are both useful in different ways. The latter classification is straightforward when the value of $C_{p,1}$ is known; however, sometimes $C_{p,1}$ cannot be defined unambiguously (e.g., when there is no regular `front row' which is perpendicular to the wind direction). The merit of using $C_{p,Nishino}$ is that this can be predicted analytically using equation \ref{cp_nishino}.

\par \textcolor{red}{It should be noted that figure \ref{fig:loss_factors} is for actuator discs (i.e., ideal turbines). Real turbines will experience additional power losses due to practical (non-ideal) turbine design.}

\par As demonstrated earlier in figures \ref{fig:LES_cp}, \ref{fig:LES_turbinelossfactor} and \ref{fig:LES_lossfactorratio}, the strength of atmospheric response alters \textcolor{blue}{wind farm performance}. The importance of array density and turbine layout vary with the large-scale atmospheric conditions. \textcolor{blue}{The results shown in appendix \ref{section:UM_farm_data}} suggests that the value of $\zeta$ tends to decrease as the wind farm size increases (at least within the range between 10km to 30km; note that the values of $\zeta$ reported in \citet{Patel2021} were for a fixed wind farm size of 20km). These results seem to suggest that larger wind farms will be less sensitive to turbine-scale wake effects and more sensitive to farm-scale losses. This could have significant implications for the design of future wind farms.

\par It is also worth noting that similar trends of large wind farm aerodynamics have already been reported in the literature, e.g., in the wind farm LES performed by \cite{Wu2017} \textcolor{blue}{in which they also used a relatively low surface roughness length of 0.05m.} They performed LES of finite and infinite wind farms under a weak and strong free-atmosphere stratification, using a rotational actuator disc model (representing Vestas V-80 2MW turbine) with an aligned and a staggered turbine layout with the same array density. The varying free-atmosphere stratifications changes the strength of atmospheric response, which affects the wind farm blockage. Figure 11 in \cite{Wu2017} shows the power production of different turbine rows for the different wind farms. For both atmospheric conditions, the layout of the infinite wind farm did not change the farm power. This agrees with our finding that the power output of infinite wind farms are insensitive to turbine interactions. The stronger free-atmosphere stratification induced a larger pressure gradient across the farm which implies a larger $\zeta$ value. The power output of the finite farm under the strong free-atmosphere stratification was more sensitive to the turbine layout. Under a weak stratification the layout had a smaller impact on farm power. This is further confirmation that the large-scale atmospheric response changes the importance of turbine-scale interactions. \textcolor{blue}{Other LES studies have also generally found that the power output of finite wind farms \citep[e.g.,][]{Porte-Agel2013,Stevens2014,Archer2013} is more sensitive to turbine layout than infinite farms \citep[e.g.,][]{yang2012,yang2014,Abkar2013}. However, these previous studies used surface roughness lengths typical of rough onshore locations rather than offshore, meaning that the ratio of total turbine drag to surface drag (equation \ref{ratioturbinesurfacedrag}) was smaller than the present study.}

\par The impact of turbine layout on wind farm performance is typically investigated by reporting the normalised turbine power, where the turbine power is normalised by either the power of a turbine in the first row or that of a standalone turbine. \textcolor{red}{The normalised power is often used as a measure of turbine-wake interactions within the wind farm.} However, this could be misleading for a large wind farm where the power is also reduced due to the ABL response to the farm resistance. The turbine-scale loss factor, $\Pi_T$, gives the power losses due to wake interactions separated from farm-scale effects.
Our results suggest that the power losses in large farms with the same layout can change with the strength of large-scale atmospheric response. Hence, the relative performance of different turbine layouts can also vary with the strength of atmospheric response. Layouts which produce more power with a specific large-scale atmospheric response may not perform as well under different responses.

\par A limitation of this study is that the wind farm LES consider only neutrally stratified atmospheric boundary layers. One uncertainty is how different stratifications would affect wake recovery within large wind farms, which could affect the $\overline{C_T^*}$ value for a given turbine layout. However, since the two-scale momentum theory has been derived from the principle of momentum conservation which is valid for all atmospheric conditions, we expect that the trends found in this study (on how the importance of turbine-scale interactions changes with the strength of the atmospheric response) would be observed generally. The study also used a fixed turbine operating condition given by $C_T'=1.33$. This is typical for the current wind turbines operating below their rated wind speed (at which the farm blockage effects are most significant). Therefore the results of this study give the trends for large farms if the operating conditions are similar to those currently used. As discussed earlier, the trends found in this study have also been observed in LES of finite wind farms with different atmospheric conditions and a different turbine model, supporting the generality of our findings.  

\section{Conclusions}\label{conclusions}

\par In this study we performed a combined theoretical and computational analysis of large wind farms, using the two-scale momentum theory \citep{Nishino2020} and new LES of infinitely large wind farms with different turbine layouts and wind directions. To consider a range of wind directions, we used a new implementation of the actuator disc model which is not aligned with the structured grid. Firstly, we validated the LES against the results published by \citet{Calaf2010} and found an excellent agreement. We also confirmed that, when the correction factor proposed by \citet{Shapiro2019} is applied, the average internal turbine thrust coefficient $\overline{C_T^*}$ (representing the turbine thrust relative to the square of the average wind speed across the wind farm layer) is insensitive to the grid resolution. We then performed 50 wind farm LES with different turbine spacings and wind directions. $\overline{C_T^*}$ was found to be a strong function of wind direction with a weaker dependence on turbine spacings.  

\par This study adopted an approach to adjust the average power coefficient $\overline{C_p}$ of an infinitely large wind farm to estimate the power of large but finite-sized farms with the same layout, following the concept of the two-scale momentum theory. \textcolor{red}{A full validation of this approach is currently infeasible as it would require a large set of finite-size wind farm LES, which will be the subject of future studies. However, we have provided detailed theoretical arguments on why we can estimate the power of large finite-size farms from the results of infinitely large farm LES.} For infinitely large farms, $\overline{C_p}$ was found to be a strong function of the array density and agree remarkably well with the analytical model (equation \ref{cp_nishino}). The power produced by large finite-sized farms was found to depend on both the array density and turbine layout (including the effect of wind direction), although the analytical model still provides an approximate upper limit of the power (for a given array density). These results confirm that, to model large but finite-sized wind farms it is important to consider both the effect of array density and the turbine layout.

\par The impact of array density and turbine layout on farm power varies with the strength of atmospheric response or the `wind extractability'. \textcolor{blue}{The analytical model seems to predict very well the impact of array density for a given farm-scale atmospheric response. As such, }we propose a new classification of power losses in large wind farms by introducing new metrics, namely a turbine-scale loss factor $\Pi_T$, a farm-scale loss factor $\Pi_F$ and a total loss factor $\Pi$. The turbine-scale loss factor describes the power losses due to turbine-scale wake interactions only (which is smaller than what is commonly called `wake losses'). Importantly, the turbine-scale loss factor varies with the strength of large-scale atmospheric response. As an example, $\Pi_T$ varied from 0.13 to 0.22 for a given layout across a realistic range of the extractability factor (from $\zeta=5$ to $25$ in this study). Although further studies are required to better quantify the extractability factor, the results obtained to date suggest that as wind farms get larger the proportion of power losses due to turbine-wake interactions will decrease. This is because (1) having a better turbine layout (with less wake interactions and higher $C_T^*$) does not substantially increase the momentum available to the farm site when $\zeta$ is small (figure \ref{fig:drag_composition}) and (2) $\zeta$ seems to decrease as the wind farm size increases. The farm-scale loss factor describes the power loss due to the atmospheric response to the array density (which is larger than what is often called `farm blockage losses'). $\Pi_F$ was typically more than twice as large as $\Pi_T$ for the 50 offshore wind farms considered in this study with a realistic range of atmospheric responses. This suggests that farm-scale flow effects have a greater impact than turbine-scale flow effects on the performance of large offshore wind farms.

\backsection[Acknowledgements]{The first author (AK) acknowledges the NERC-Oxford Doctoral Training Partnership in Environmental Research (NE/S007474/1) for funding and training. We acknowledge use of the Monsoon2 system, a collaborative facility supplied under the Joint Weather and Climate Research Programme, a strategic partnership between the Met Office and the Natural Environment Research Council. We also thank Dr Fran\c{c}ois-Xavier Briol for helpful discussions and assistance with the design of numerical experiments (figure \ref{fig:experiments}b) \textcolor{blue}{and Joseph Ashton for assistance with the analysis of NWP data shown in appendix \ref{section:UM_farm_data}.}}

\backsection[Funding]{This work was supported by the Natural Environmental Research Council (NERC) award NE/S007474/1.}

\backsection[Declaration of Interests]{The authors report no conflict of interest.}

\backsection[Data availability statement]{The data that support the findings of this study are openly available at https://github.com/AndrewKirby2/wind-farm-modelling. This includes the wind farm LES results and the code used to calculate the data for figures \ref{fig:LES_cp}, \ref{fig:LES_turbinelossfactor} and \ref{fig:LES_lossfactorratio}.}

\backsection[Author ORCID]{A. Kirby, https://orcid.org/0000-0001-8389-1619; T. Nishino, https://orcid.org/0000-0001-6306-7702.}

\backsection[Author contributions]{T.N. derived the theory, A.K. and T.D.D. performed the simulations, A.K. and T.N. analysed the data. A.K. wrote the paper with corrections from T.N. and T.D.D.}

\appendix

\section{Sensitivity of results to nominal farm-layer height $H_F$}\label{hf_sensitivity}

The height of the farm-layer, $H_F$, can be defined in several ways. This study uses the definition $H_F=2.5H_{hub}$ where $H_{hub}$ is the turbine hub height. Differences in the value of $H_F$ will change the values of $U_F$, $U_{F0}$, $\overline{C_T^*}$ and $C_{f0}$ of the LES results presented in section \ref{results}. However, as noted by \citet{Nishino2020}, the conservation of momentum arguments used to derive equation \ref{windfarmmomentum} and to eventually predict $C_p$ from equation \ref{cp}, are valid irrespective of the $H_F$ value, provided that the same $H_F$ value is used in both `internal' and `external' sub-problems. The question here is how much the results of `internal' sub-problem (i.e., LES results in this study) and `external' sub-problem (i.e., the value of $\zeta$ obtained from NWP simulations) change with the choice of $H_F$. Although the `external' sub-problem is outside the scope of this paper, \citet{Patel2021} reported that the sensitivity of their NWP results to the choice of $H_F$ was minor. In the following, we present how the sensitivity of the LES results to the choice of $H_F$ eventually affects the value of $C_p$ obtained (assuming that $\zeta$ is not affected by $H_F$).

\begin{figure*}
\centering
\includegraphics[width=\textwidth]{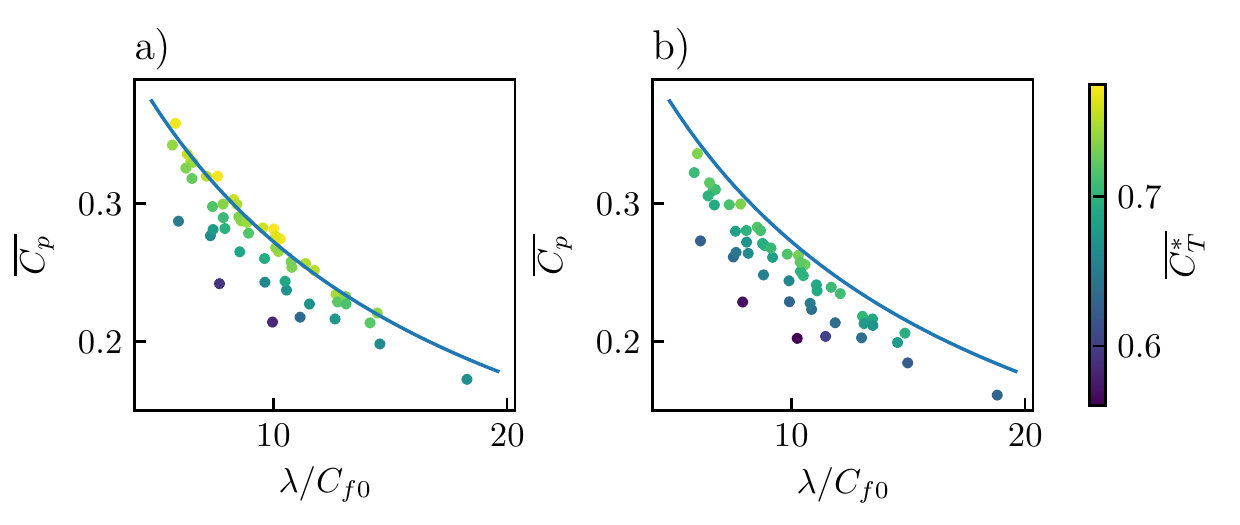}
\caption{LES and theoretical results for $\overline{C_p}$ with $\zeta=25$ for a) $H_F$=250m and b) $H_F$=300m.}
\label{fig:hf_sensitivity}
\end{figure*}

To investigate the sensitivity of the results to $H_F$, we repeat the analysis with $\zeta=25$ using $H_F=3H_{hub}$. Figure \ref{fig:hf_sensitivity} compares the  $\overline{C_p}$ values obtained for $H_F=2.5H_{hub}$ and $3H_{hub}$. Changing the value of $H_F$ has no effect on the theoretical predictions of $C_p$ for a given $\zeta$. Increasing $H_F$ decreases the LES-based results of $\overline{C_p}$ for each farm because the reference velocity $U_{F0}$ slightly increases. The increase in $U_{F0}$ is the same for all wind farm cases. Therefore, increasing $H_F$ to $3H_{hub}$ reduces the $\overline{C_p}$ of all wind farm cases by the same factor of $1.043$ in this study. The absolute values of $\overline{C_p}$ are affected by the value of $H_F$ but the trends reported in this study remain unchanged. The $\overline{C_T^*}$ values obtained from the LES are also reduced when the farm-layer height is increased. This is because the reference velocity $U_F$ slightly increases. Nevertheless, the two-scale momentum theory still captures the trend well and can provide an upper bound estimate to wind farm performance.

\cite{Nishino2020} proposed a definition for $H_F$ based on the undisturbed velocity profile $\overline{U_0(z)}$. $H_F$ is defined as the farm-layer height with which the farm-layer average of the undisturbed velocity is equal to that of the rotor average, i.e.,

\begin{equation}
    \frac{\int_0^{H_F}\overline{U_0}dz}{H_F}=\frac{\int\overline{U_0}dA}{A}.
    \label{hf_definition}
\end{equation}

\begin{figure*}
\centering
\includegraphics[width=0.5\textwidth]{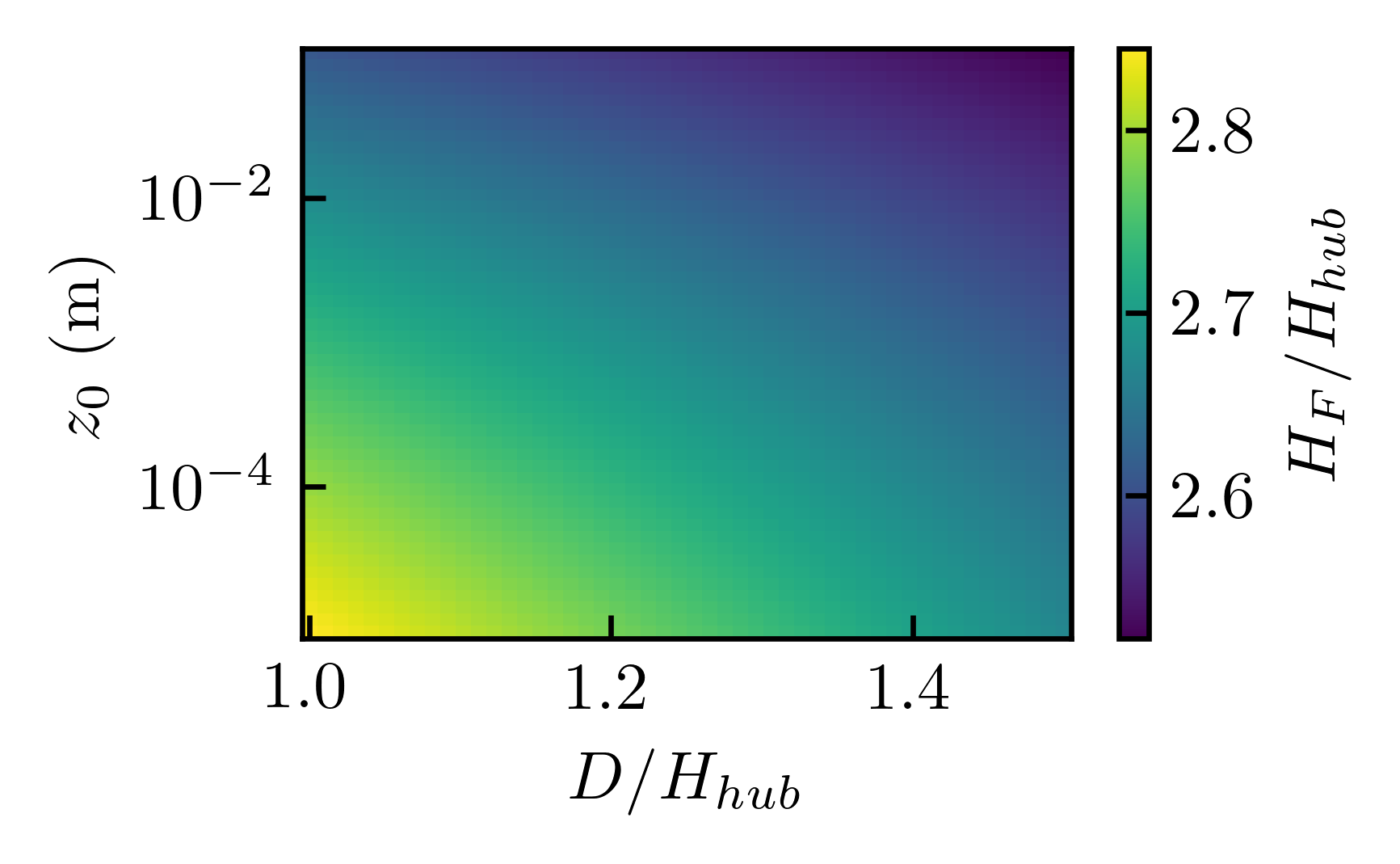}
\caption{Variation of normalised wind farm-layer height $H_F/H_{hub}$ with surface roughness length $z_0$ and turbine design parameter $D/H_{hub}$ for a log law wind profile.}
\label{fig:hf_values}
\end{figure*}

\noindent Using this definition, the exact value of $H_F$ will vary with turbine design and (undisturbed) ABL profile. Figure \ref{fig:hf_values} shows the value of $H_F$ according to \ref{hf_definition} for a wide range of velocity profiles and turbine designs. We assume that the velocity profile is given by the logarithmic law for a wide range of surface roughness lengths that includes onshore and offshore locations. For the turbine design, we assume that the distance between the surface and bottom of the rotor is fixed at 50m. The range of $D/H_{hub}$ used corresponds to turbine diameters from 100m up to 300m. Across a wide range of turbine design and velocity profiles the value of $H_F$ obtained from equation \ref{hf_definition} only varies between $2.5H_{hub}$ and $3H_{hub}$.

\par The average turbine power coefficient $\overline{C_p}$ is changed only slightly when $H_F$ is increased from $2.5H_{hub}$ to $3H_{hub}$. The trends reported in this study are unaffected by a change in $H_F$. $H_F$ is relatively insensitive to velocity profiles and turbine design and is expected to be in the range $2.5H_{hub}$ to $3H_{hub}$. \textcolor{blue}{Therefore, the results of the present study are insensitive to the exact value of $H_F$.}

\FloatBarrier

\section{\textcolor{blue}{Effect of wind farm size on the wind extractability factor $\zeta$}}\label{section:UM_farm_data}

\par
\begin{figure*}
\centering
\includegraphics[width=0.75\textwidth]{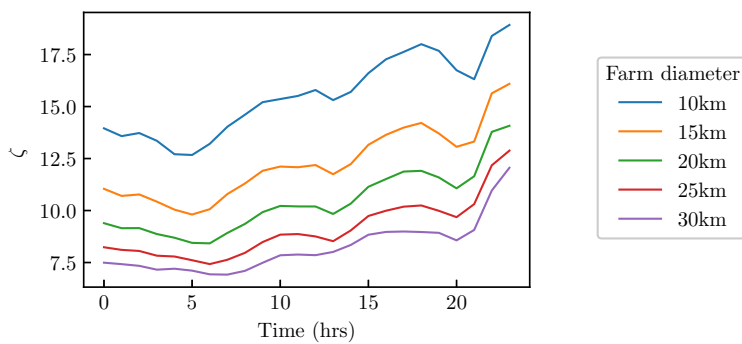}
\caption{Variation of $\zeta$ throughout a 24 hour period (corresponding to Case B in figure 9 of \citet{Patel2021}) for different wind farm diameters.}
\label{fig:UM_zeta}
\end{figure*}

 \textcolor{blue}{To explore the effect of wind farm size on the wind extractability factor $\zeta$, we performed additional twin NWP simulations using the same methodology as \citet{Patel2021}. Within the simulations, the diameter of a hypothetical circular wind farm (located off the east coast of Scotland) was varied from 10km to 30km. Example results for a 24-hour period with a relatively constant wind speed across the farm site (corresponding to Case B of figure 9 of \citet{Patel2021}) are shown in figure \ref{fig:UM_zeta}. It can be seen that the value of $\zeta$ tends to decrease with increasing the wind farm size for all the time. Further simulations are required in future studies to derive a correction function for the effect of wind farm size on $\zeta$, but our preliminary results suggest that an exponential relationship may exist between the wind farm size and $\zeta$. Such an exponential correction would satisfy the expected asymptotic behaviour of $\zeta$ (i.e., $\zeta$ approaches to +$\infty$ and 0 as the farm size approaches to 0 and +$\infty$, respectively). Figure \ref{fig:UM_zeta} also shows that $\zeta$ varies significantly with time due to changing atmospheric conditions. These preliminary results show that $\zeta$ depends both on wind farm size and atmospheric conditions.}

\FloatBarrier

\bibliographystyle{jfm}
\bibliography{references.bib}


\end{document}